\documentclass[twocolumn,epjc3]{svjour3}  
\usepackage{amsmath,amssymb,amsfonts}
\usepackage{graphicx,graphics}
\usepackage{graphics}
\usepackage{url}
\usepackage{color}
\usepackage{mathtools}
\usepackage[utf8]{inputenc}

\journalname{Eur. Phys. J. C}

\usepackage[utf8]{inputenc}

\title{Can we fit our nuclear PDFs with the high-$x$ CLAS data?}

\author{Hannu Paukkunen$^{1,2,}$\thanksref{e1} \and Pia Zurita$^{3,}$\thanksref{e2}}
\date{February 2020}

\thankstext{e1}{e-mail: hannu.paukkunen@jyu.fi}
\thankstext{e2}{e-mail: maria.zurita@ur.de}

\begin{document}

\institute{University of Jyvaskyla, Department of Physics, P.O. Box 35, FI-40014 University of Jyvaskyla, Finland \and
Helsinki Institute of Physics, P.O. Box 64, FI-00014 University of Helsinki, Finland \and 
Institut f\"ur Theoretische Physik, Universit\"at Regensburg, 93040 Regensburg, Germany}

\maketitle

\abstract{
Nuclear parton distribution functions (nuclear PDFs) are non-perturbative objects that encode the 
partonic behaviour of bound nucleons. To avoid potential higher-twist contributions, the data probing the high-$x$ end of nuclear PDFs are sometimes left out from the global extractions despite their potential to constrain the fit parameters. In the present work we focus on the kinematic corner covered by the new high-$x$ data measured by the CLAS/JLab collaboration. By using the Hessian re-weighting technique, we are able to quantitatively test the compatibility of these data with globally analyzed nuclear PDFs and explore the expected impact on the valence-quark distributions at high $x$. We find that the data are in a good agreement with the EPPS16 and nCTEQ15 nuclear PDFs whereas they disagree with TuJu19. The implications on flavour separation, higher-twist contributions and models of EMC effect are discussed. 
}


\section{Introduction}

The nuclear parton distribution functions (nuclear PDFs) \cite{Paukkunen:2018kmm,Paukkunen:2017bbm} quantifying the structure of quarks and gluons in bound nucleons constitute an indispensable ingredient in precision calculations for processes at high interaction scales $Q^2 \gg \Lambda^2_{\rm QCD}$ in high-energy colliders like the Large Hadron Collider (LHC). Based on the collinear factorization theorem \cite{Collins:1989gx}, nuclear PDFs are believed to be process independent and the scale dependence to follow the usual linear DGLAP evolution \cite{Dokshitzer:1977sg,Gribov:1972ri,Gribov:1972rt,Altarelli:1977zs}. These assumptions have been observed to be consistent with experimental data ranging from deeply inelastic scattering (DIS) to heavy-ion collisions. For example, although in high-energy lead-lead collisions there is evidence for the formation of a state that effectively behaves as a strongly-interacting liquid, the electroweak observables \cite{Aad:2015lcb,Aad:2014bha,Acharya:2017wpf,Chatrchyan:2014csa} are consistent with the nuclear PDF predictions. Moreover, the linear DGLAP evolution in proton-lead collisions has been verified down to $x \sim 10^{-5}$ at low $Q^2$ through heavy-quark production \cite{Eskola:2019bgf}, with no evidence of a breakdown.

It is well known that the PDFs are best constrained through DIS experiments. Indeed, thanks to the HERA data \cite{Abramowicz:2015mha}, the free-proton PDFs are quite well determined in a wide kinematic window. The regimes where one has to rely on extrapolations are limited to the very small $x$ ($x<10^{-5}$) and the high-$x$ regions. The former, due to not having been explored in electron-proton experiments; the latter due to the imposition of kinematic cuts on $Q^2$ and the final-state invariant mass $W$ to avoid potentially large higher-twist contributions such as target mass corrections.\footnote{These cuts are routinely applied in proton PDF fits, though their relaxation has been explored, e.g. by the CTEQ-JLab collaboration \cite{Accardi:2016qay}.} However, there has not yet been an experiment equivalent to HERA with nuclear beams -- only fixed-target DIS data spanning a rather limited region of the kinematic space (though covering a variety of nuclei) are available. With no ``nuclear HERA data" the nuclear PDFs still suffer from large uncertainties and e.g. the flavour separation is only poorly known. Given the fact that high-energy nuclear DIS at the Electron-Ion Collider (EIC) \cite{Accardi:2012qut} or at the planned LHeC/FCC-eh \cite{AbelleiraFernandez:2012cc} are at least a decade away, the community has generally sought to improve the situation by using the LHC proton-lead data as new constraints in the global analyses. 

An additional possibility is to aim at a more complete use of the already available high-$x$ DIS data. Imitating the typical free-proton fits some of the nuclear-PDF analyses also set stringent cuts on $Q^2$ and $W$. Given the low center-of-mass energies of the available fixed-target data, a significant fraction of the data get easily cut away. Lowering the minimum value of $Q^{2}$ one reaches lower in $x$, while lowering the cut in the final-state invariant mass $W$ the high-$x$ low-$Q^{2}$ data enter the fits. In the present paper we concentrate on this latter regime by studying the compatibility and impact of the very precise DIS data measured recently by the CLAS collaboration \cite{Schmookler:2019nvf} by using recent sets of nuclear PDFs at a next-to-leading order (NLO) accuracy. The data were taken in the high-$x$ region ($0.2<x<0.6$) where the so-called EMC effect \cite{Arneodo:1992wf,Malace:2014uea} occurs. On one hand, the approach based on nuclear PDFs is phenomenological in the sense that one does not address the underlying microscopic dynamics of the nuclear effects: Based solely on the framework of QCD and collinear factorization, the predictions from nuclear PDFs aim to be \emph{model independent}. On the other hand, in the same $x$ range but at higher $Q^2$ there are other lepton-nucleus DIS data e.g. from SLAC/NMC collaborations \cite{Gomez:1993ri,Arneodo:1996rv,Amaudruz:1995tq} and also neutrino-nucleus DIS data from e.g. the CHORUS collaboration \cite{Onengut:2005kv}. In addition, recent CMS dijet data \cite{Sirunyan:2018qel} have been found to be sensitive to the valence-quark EMC effect. Since the $Q^2$ dependence of the EMC effect in global analyses of nuclear PDFs is fully dynamical, dictated by the DGLAP evolution, an ability (or disability) to describe all these data with an universal initial condition for the $Q^2$ dependence will (i) quantitatively address the importance of possible higher-twist $\sim Q^{-2n}$ contributions and (ii) place restrictions on the possible origin of the EMC effect. In particular, the current nuclear PDF analyses all assume that the strength of the nuclear effects scale as a function of nuclear mass number $A$, although e.g. models for the EMC effect based on short-range correlations \cite{Hen:2016kwk} would suggest also an isospin dependence. The current global fits of nuclear PDFs, however, have not found support for an existence of such a component. This only makes the new JLab/CLAS data more welcome and constitute an interesting test bench for the nuclear PDFs. 

The rest of the document is organised as follows. In Sec.~\ref{section:framework} we introduce the framework of our study, including the nuclear PDF sets, details regarding the calculation, target mass corrections and the method employed. In Sec. \ref{section:results} we discuss our results including the potential of the data to further improve our knowledge of the valence distributions at high{\textcolor{magenta} {-}}$x$. Finally we summarise our results in Sec. \ref{section:summary}.  

\section{Framework}\label{section:framework}

\subsection{Nuclear PDFs at high $x$} \label{subsection:HixPDFs}

\begin{figure}[htb!]
\centering
\includegraphics[width=1.0\linewidth]{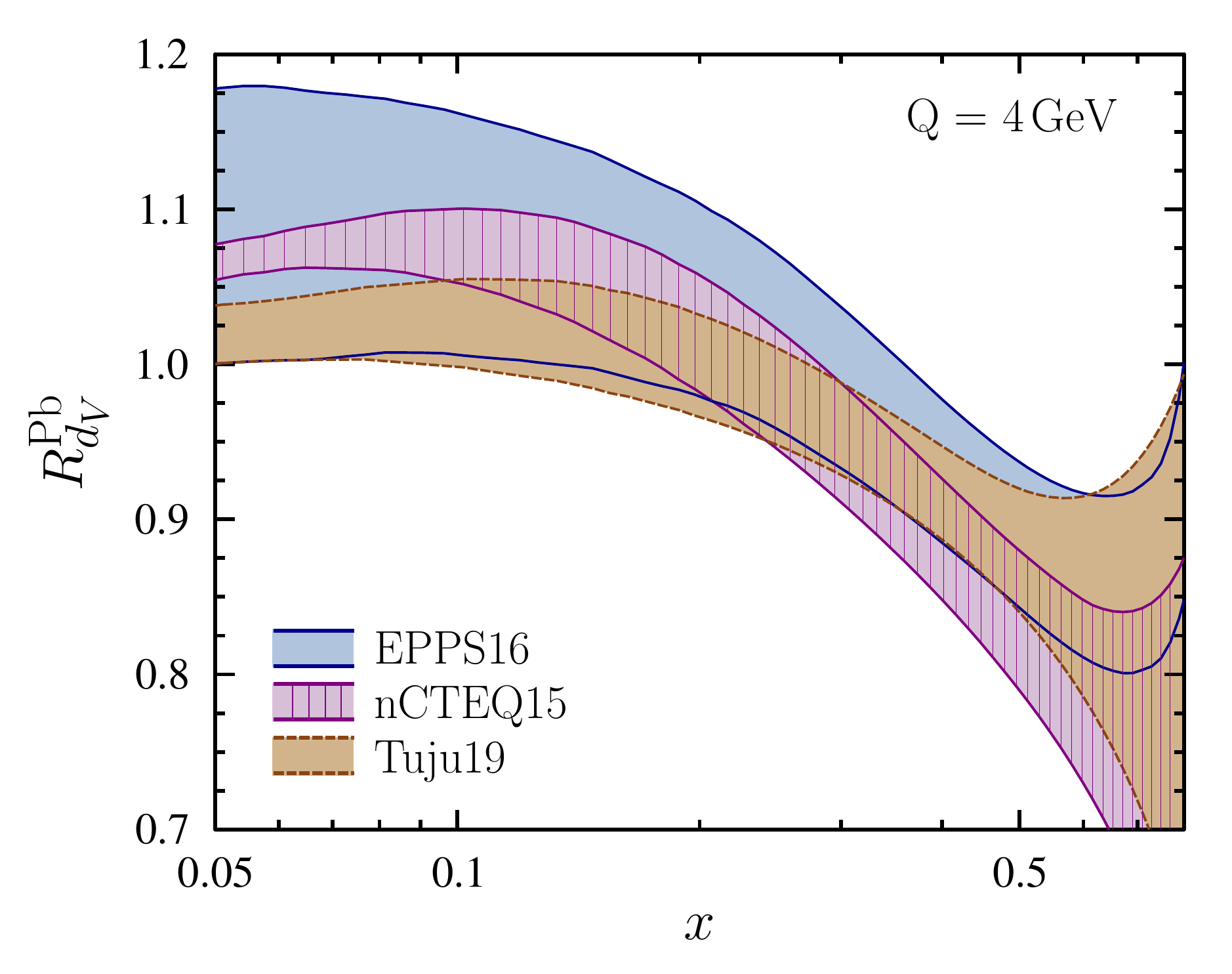}
\includegraphics[width=1.0\linewidth]{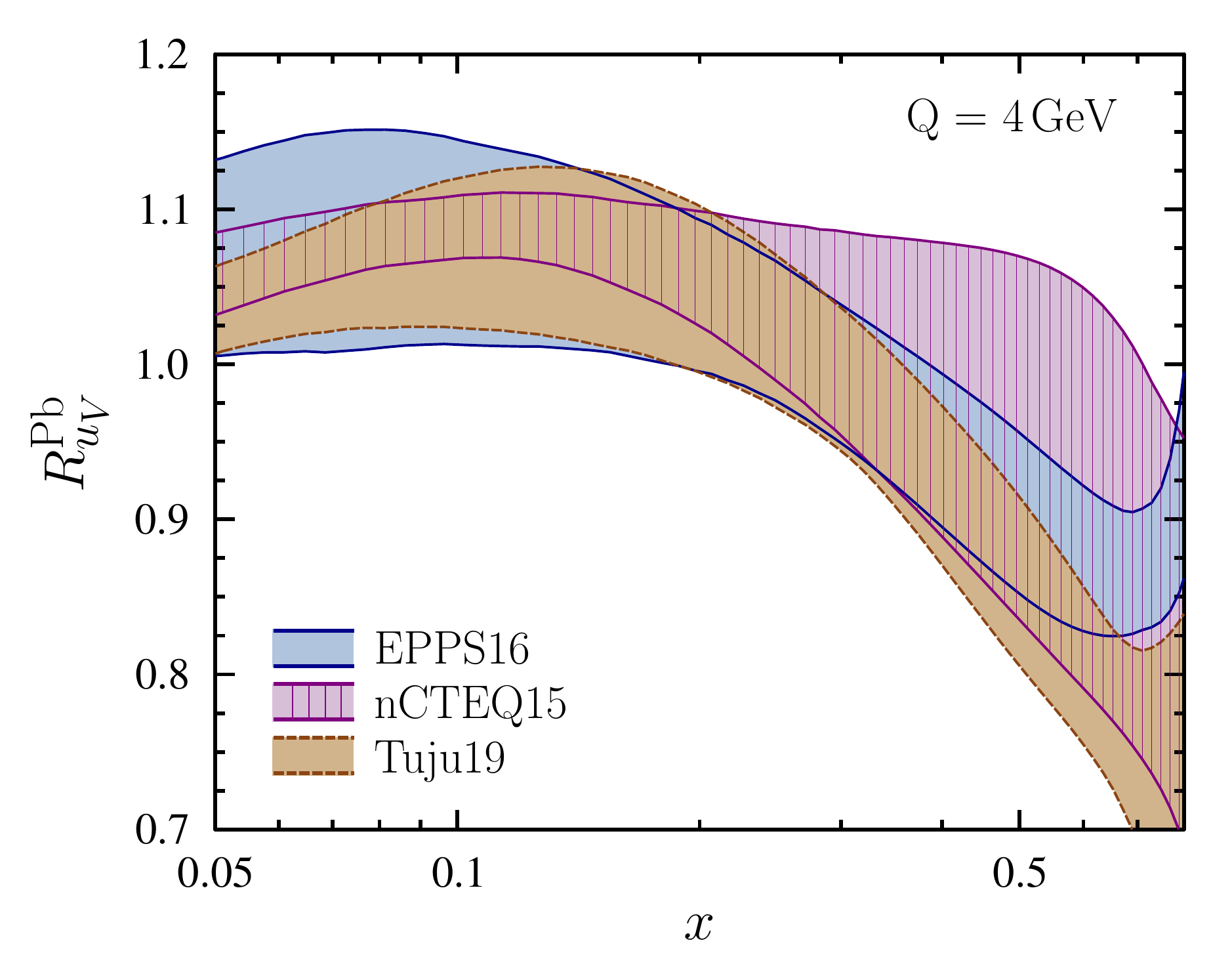} \\
\caption{Comparison of the valence-quark nuclear modifications as encoded in the EPPS16 (blue bands), nCTEQ15 (purple bands with hatching) and TuJu19 (brown bands) parametrizations at $Q = 4 \, {\rm GeV}$.}
\label{fig:valenceall}
\end{figure} 

In our present study, we will utilize three modern sets of nuclear PDFs: EPPS16 \cite{Eskola:2016oht}, nCTEQ15 \cite{Kovarik:2015cma} and TuJu19 \cite{Walt:2019slu}. All these three sets involve the valence-quark flavour separation and thereby better reflect the prevailing uncertainties at large $x$. We refrain here from using sets with no flavour separation (e.g. EPS09 \cite{Eskola:2009uj}, DSSZ12 \cite{deFlorian:2011fp}, KA15 \cite{Khanpour:2016pph}, nNNPDF1.0 \cite{AbdulKhalek:2019mzd}). As it is customary in the case of nuclear PDFs, we will discuss the behaviour of the nuclear valence distributions in terms of certain ratios which better reflect the relevant features of nuclear PDFs. We define here the valence-quark nuclear modifications $R_{u_V}^A(x,Q^2)$ and $R_{d_V}^A(x,Q^2)$, as the total up/down valence distribution in a nucleus $A$ with $Z$ protons and $N$ neutrons, divided by the same distribution but with no nuclear effects in the PDFs,
\begin{align}
R_{u_V}^{\rm Pb}(x,Q^2) & \equiv \frac{u_V^A(x,Q^2)}{Z u_V^{\rm p}(x,Q^2) + N d_V^{\rm p}(x,Q^2)} \,, \label{eq:Ruvtot1}  \\
R_{d_V}^{\rm Pb}(x,Q^2) & \equiv 
\frac{d_V^A(x,Q^2)}
{Z d_V^{\rm p}(x,Q^2) + N u_V^{\rm p}(x,Q^2)} 
\,. \label{eq:Rdvtot1}
\end{align}
Here $u_V^{\rm p}(x,Q^2)$ and $d_V^{\rm p}(x,Q^2)$ denote the free-proton valence-quark PDFs. When forming the ratios the proton PDF used is always the one taken as baseline in the corresponding nuclear-PDFs analysis (e.g. CT14NLO for EPPS16). These ratios are plotted in Fig.~\ref{fig:valenceall} for the lead (Pb) nucleus. In general, there seems to be a fair agreement between different parametrizations, though the shapes and widths of the uncertainty bands differ from each other. In particular, the nCTEQ15 uncertainty for $R_{u_V}^{\rm Pb}(x,Q^2)$ is clearly larger than those of EPPS16 and TuJu19. This is presumably due to the facts that the nCTEQ15 analysis (i) used isoscalar DIS data which skews the flavour separation and (ii) did not include neutrino DIS data which e.g. in the EPPS16 analysis clearly improved the flavour separation -- presumably so also in TuJu19. Indeed, the combination of PDFs probed in neutral-current DIS is of the form (suppressing the $x$ and $Q^2$ arguments),
\begin{align}
4u_V^A + d_V^A & = 4 \left[{Z R_{u_V}^{p/A} u_V^{\rm p} + N R_{d_V}^{p/A}d_V^{\rm p}}\right]
\label{eq:NCDIS}
\\ 
 & \hspace{0.3cm} +
\left[{Z R_{d_V}^{p/A}d_V^{\rm p} + N R_{u_V}^{p/A} u_V^{\rm p}} \right] \nonumber \\
& \propto
R_{u_V}^{p/A} + R_{d_V}^{p/A} \times 
\frac{d_V^{\rm p}}{u_V^{\rm p}} \frac{Z  + 4N }{4Z  + N} \nonumber \\
& \xrightarrow{N=Z}
R_{u_V}^{p/A} + R_{d_V}^{p/A} \times \frac{d_V^{\rm p}}{u_V^{\rm p}}
 \nonumber \,,
\end{align}
where $R_{u_V}^{p/A}$ and $R_{d_V}^{p/A}$ are nuclear modifications of the bound protons (these are what nCTEQ15, EPPS16 and TuJu19 effectively parametrized). Since the valence up is roughly twice-thrice the valence down at high-$x$, we have $d_V^{\rm p}/u_V^{\rm p} \ll 1$ at large $x$, and it is clear that the relative weight of $R_{u_V}^{p/A}$ in the equation above is larger and 
therefore better constrained by the fit. Since the ratio $d_V^{\rm p}/u_V^{\rm p}$ is not constant as a function of $x$, the linear combination of $R_{u_V}^{p/A}$ and $R_{d_V}^{p/A}$ does not remain constant and through the assumed form of the parametrization one can constrain them separately to some extent even with isoscalar nuclei ($N=Z=A/2$). On the other hand we can write Eqs.~(\ref{eq:Ruvtot1}) and (\ref{eq:Rdvtot1}) also as 
\begin{align}
R_{u_V}^{\rm Pb} & =
\frac{Z R_{u_V}^{\rm p/Pb} u_V^{\rm p} + N R_{d_V}^{\rm p/Pb} d_V^{\rm p}}{Z u_V^{\rm p} + N d_V^{\rm p}}
\,, \label{eq:Ruvtot2} \\
R_{d_V}^{\rm Pb} & =
\frac{Z R_{d_V}^{\rm p/Pb} d_V^{\rm p} + N R_{u_V}^{\rm p/Pb} u_V^{\rm p}} {Z d_V^{\rm p} + N u_V^{\rm p}} \label{eq:Rdvtot2}
\,,
\end{align}
and because the better constrained component $R_{u_V}^{\rm p/Pb}$ has a larger weight in $R_{d_V}^{\rm Pb}$ ($N=126$ and $Z=82$ for $^{208}$Pb) it follows that $R_{d_V}^{\rm Pb}$ is better determined than $R_{u_V}^{\rm Pb}$ if only isoscalar neutral-current data is used as a constraint. This is clearly what we see in Fig.~\ref{fig:valenceall} for nCTEQ15. The valence-quark flavour separation in EPPS16 and TuJu19 is better constrained for using non-isoscalar data and neutrino-nucleus DIS. In Sect.~\ref{section:results} we will discuss how the features seen here are reflected in the predictions in physical DIS cross sections. 

\subsection{DIS cross sections and mass scheme}\label{subsection:DIS}

The theoretical predictions for the DIS cross-sections were computed at NLO accuracy using the simplified Aivazis-Collins-Olness-Tung (SACOT) variant of the general-mass variable-flavour-number scheme with the so-called $\chi$ rescaling \cite{Kramer:2000hn,Tung:2001mv}. This coincides with the scheme used in the EPPS16, nCTEQ15 and TuJu19 analyses.\footnote{According to Ref.~\cite{Forte:2010ta}, the FONLL-A scheme used in TuJu19 is equivalent with SACOT.} The choice of scheme is not particularly critical, however, given that the heavy-quark production does not play a significant role in the inclusive cross sections at $x \geq 0.2$, and that we are here mainly interested in ratios of cross sections,
\begin{equation}
\frac{d^2\sigma(A)}{dxdQ^2} \bigg / \frac{d^2\sigma({\rm D})}{dxdQ^2} \,,
\end{equation}
where D refers to deuteron. We have verified the scheme independence of our results by comparing our calculations to the ones in the Thorne-Roberts scheme \cite{Thorne:2012az}. We have also explicitly checked that the theoretical uncertainties for cross-section ratios do not have a sizable contribution from the baseline proton PDF uncertainties, as long as the same free-proton PDF that was the baseline PDF for each set of nuclear PDFs is used. In this respect the predictions are theoretically robust. For consistency, we neglect the nuclear effects in deuteron ($A=2$) in the case of EPPS16 and nCTEQ15. The TuJu19 parametrization, however, extends down to $A=2$ and we are able to address the role of deuteron nuclear effects. 

\subsection{Target-mass corrections}\label{subsection:tmc}

When approaching the large-$x$ and low-$Q^2$ limit, the DIS cross sections will eventually become sensitive to $1/Q^{2n}$-type power corrections originating from beyond-leading-twist contributions (not determined by PDFs) and finite nucleon mass. When $W$ is low, one also has to eventually consider effects such as nucleon resonances and in the case of nuclei, the fact that bound nucleons can carry more momentum than the average momentum per nucleon (i.e. $0<x_{\rm nucleon}<A$). In our calculations we account for the dominant part of target-mass corrections (TMCs) -- an effect that is particularly relevant at low $Q^{2}$ and high $x$. In the DIS limit \cite{Roberts:1990ww}, $Q^2, P\cdot q \to \infty$ ($\equiv$ massless quarks and nucleons), the usual Bjorken variable $x \equiv {Q^{2}}/{2P\cdot q}$ gives the fraction of light-cone momentum of the target nucleon ($P$) carried by the hit parton.\footnote{The variable $q$ marks the momentum of the virtual photon.} However at low/moderate virtualities this identification is no longer necessarily accurate. Instead, the parton light-cone momentum fraction is given by the so-called Nachtmann variable $\xi$:
\begin{align}
\xi=\frac{2x}{1+\sqrt{1+4x^{2}M^{2}/Q^{2}}} \, ,
\end{align}
where $M$ is the nucleon mass. The difference between $x$ and $\xi$ has to be considered in the calculation of the structure functions. In the present work we use the prescription of Ref.~\cite{Georgi:1976ve}, 
\begin{align}
F_{2}^{\rm TMC}(x,Q^{2}) & =\frac{x^{2}}{\xi^{2}(1+4x^{2}M^{2}/Q^{2})^{3/2}}F^{\rm LT}_{2}(\xi,Q^{2})  \,, \\
F_{L}^{\rm TMC}(x,Q^{2}) & =\frac{x^{2}}{\xi^{2}(1+4x^{2}M^{2}/Q^{2})^{1/2}}F^{\rm LT}_{L}(\xi,Q^{2}) \,,
\end{align}
where $F^{\rm LT}_{i}$ refer to leading-twist structure functions i.e. those calculated with PDFs. We neglect the corrections suppressed by additional powers of ${xM^2}/{Q^2}$ whose effect we have found negligible for the cross-section ratios. In practice, since
\begin{equation}
\frac{d^2\sigma(A)}{dxdQ^2} \bigg / \frac{d^2\sigma({\rm D})}{dxdQ^2} 
\approx \frac{F_2^A(x,Q^2)}{F_2^{\rm D}(x,Q^2)},
\end{equation}
to a very good approximation, the principal effect of TMCs in the cross section ratios is a shift in the probed value of the momentum fraction. We note that in the prescription used here $F_i^{\rm TMC}(x=1,Q^{2}) \neq 0$ which can be avoided in an alternative approach \cite{Accardi:2008ne}. However, as now $x < 0.6$ this is not yet an issue. For a review of TMCs corrections we refer the reader to \cite{Schienbein:2007gr}. The possible relevance of the other higher-twist contributions is addressed here by investigating the compatibility of the data with global fits of nuclear PDF constrained by data at higher $Q^2$. 

\subsection{The CLAS data}\label{subsection:data}

In this study we use the high-precision data measured by the CLAS collaboration \cite{Schmookler:2019nvf} and assess their potential in constraining the nuclear valence-quark distributions, particularly in the high-$x$ region. These data are ratios of inclusive electron-ion ($e^-A$) DIS cross-sections with respect to the same observable in electron-deuteron collisions. They cover the kinematic region $0.2<x<0.6$, with the average $Q^2$ spanning the range $1.62 < Q^2/{\rm GeV}^2 < 3.02$, and $W\geq 1.8$ GeV, which is just above the resonance region. In typical PDF fits these data would be discarded due to smallness of $Q^2$ and $W$ (e.g. nCTEQ15 requires $Q^2 > 2 \, {\rm GeV}^2$ and $W > 3.5 \, {\rm GeV}$) but e.g. in the EPPS16 analysis no separate cut on $W$ was imposed. There are 26 data points per target and four different nuclear targets: carbon (C), aluminium (Al), iron (Fe) and lead (Pb). In total the number of data points is thus $N_{\rm data} = 104$. We note that similar JLab data exist also for very light nuclei \cite{Seely:2009gt}.

\subsection{Hessian re-weighting and definition of $\chi^2$}\label{subsection:rw}

The impact study was done by means of the Hessian re-weighting technique \cite{Paukkunen:2013grz,Paukkunen:2014zia} in which the sensitivity of the data $\chi^2$ to the PDF error sets is translated into new PDF errors. If the variation remains much smaller than the global tolerance criterion $\Delta \chi^2$ the new data are not bound to have a significant impact, and vice versa. Re-weighting methods have become very popular in recent years to provide fast estimations of the consistency and impact of new data on existing PDFs, and play a key role in studies related to future experiments. For discussions of the different implementations of Hessian-PDF re-weighting, see \cite{Paukkunen:2014zia,Eskola:2019dui} and references therein.

\begin{figure*}[htb!]
\centering
\includegraphics[width=0.329\linewidth]{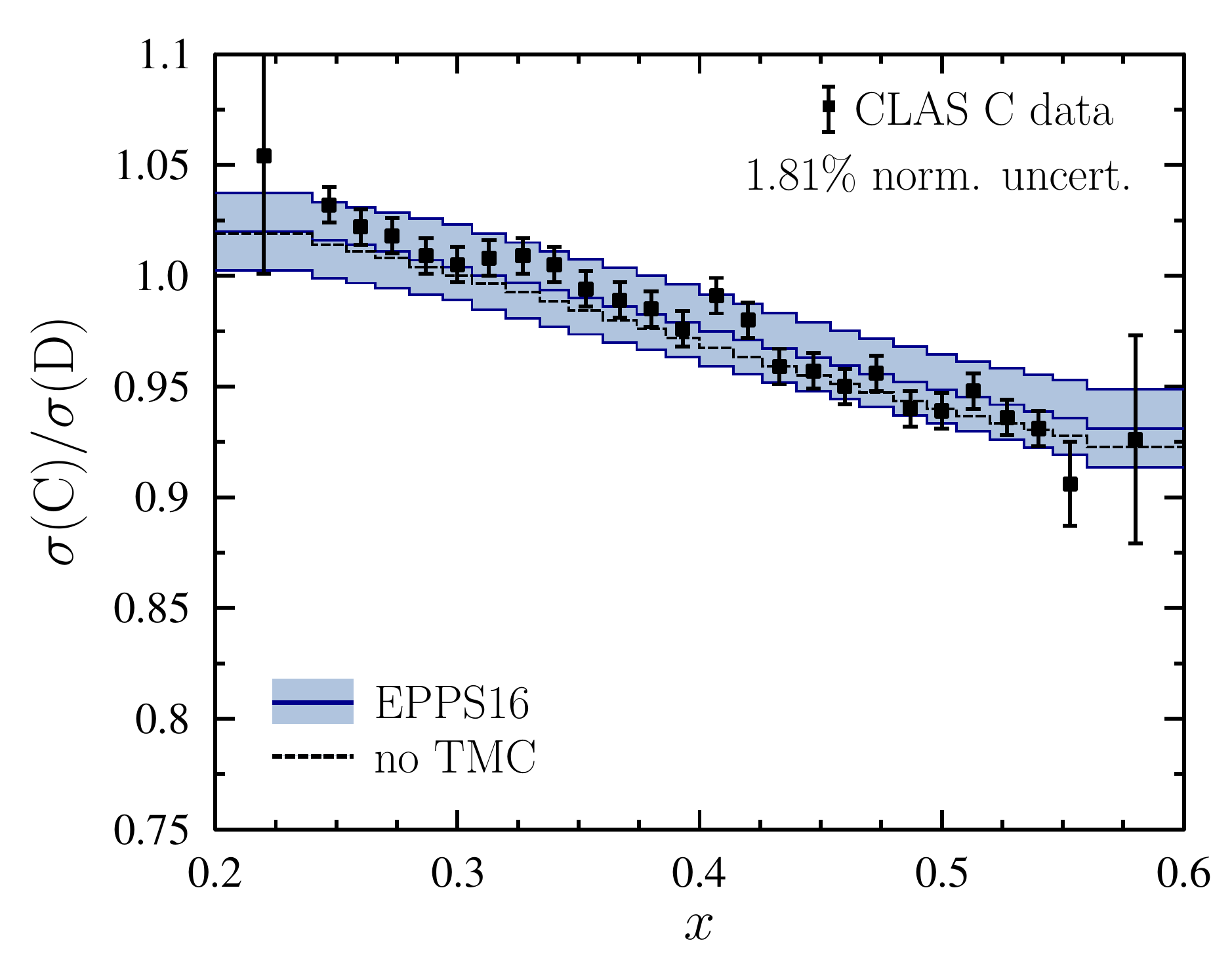}
\includegraphics[width=0.329\linewidth]{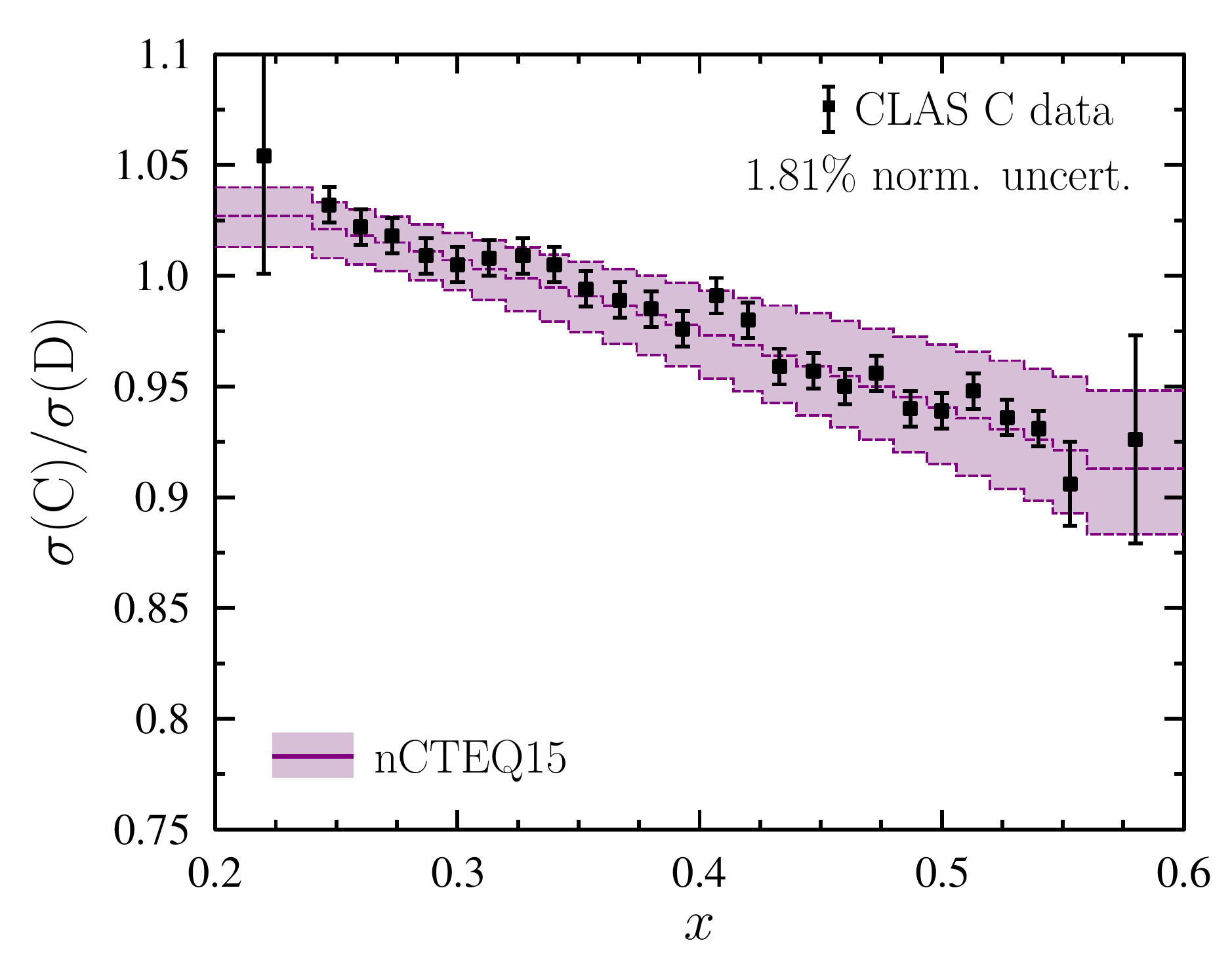}
\includegraphics[width=0.329\linewidth]{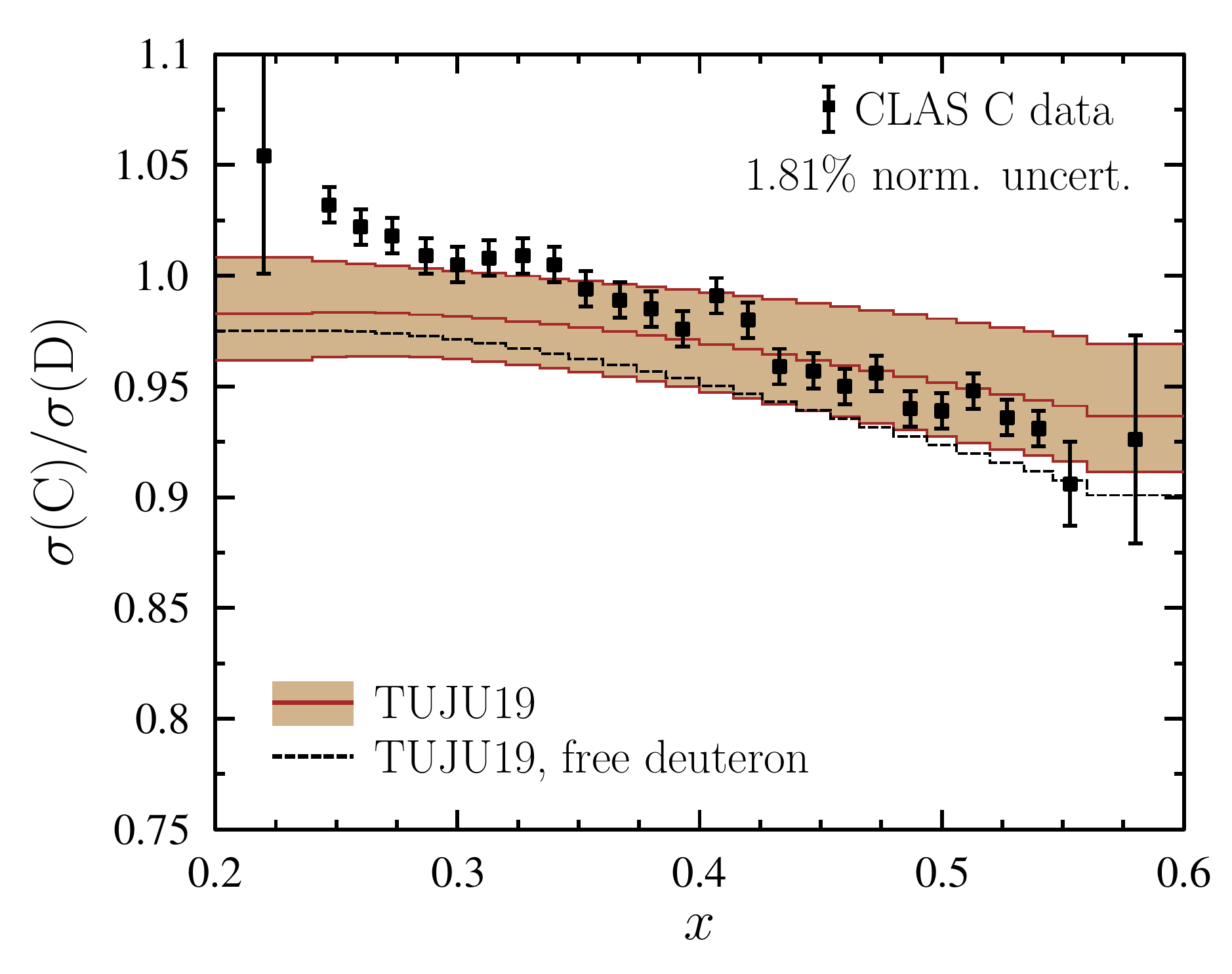}
\includegraphics[width=0.329\linewidth]{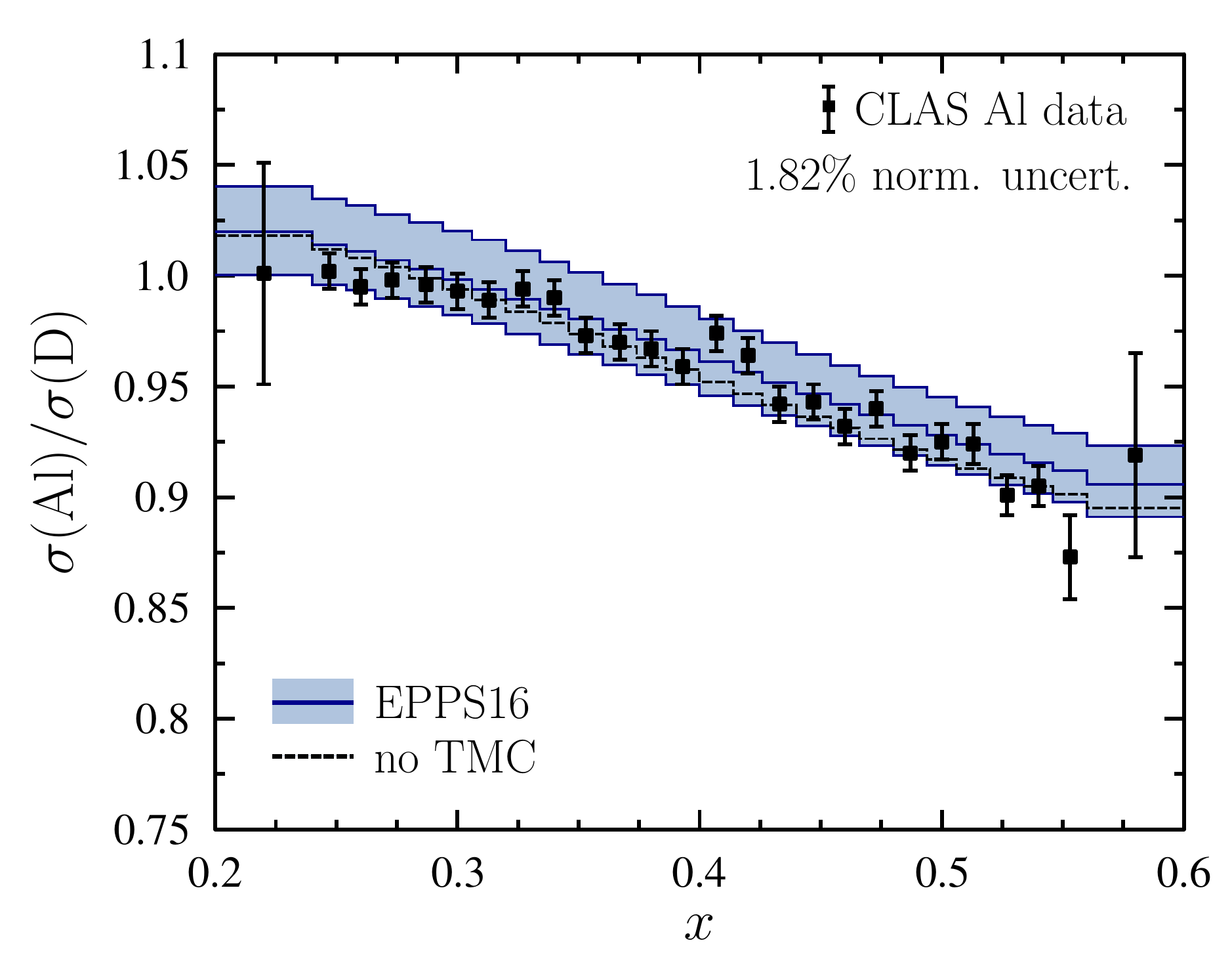}
\includegraphics[width=0.329\linewidth]{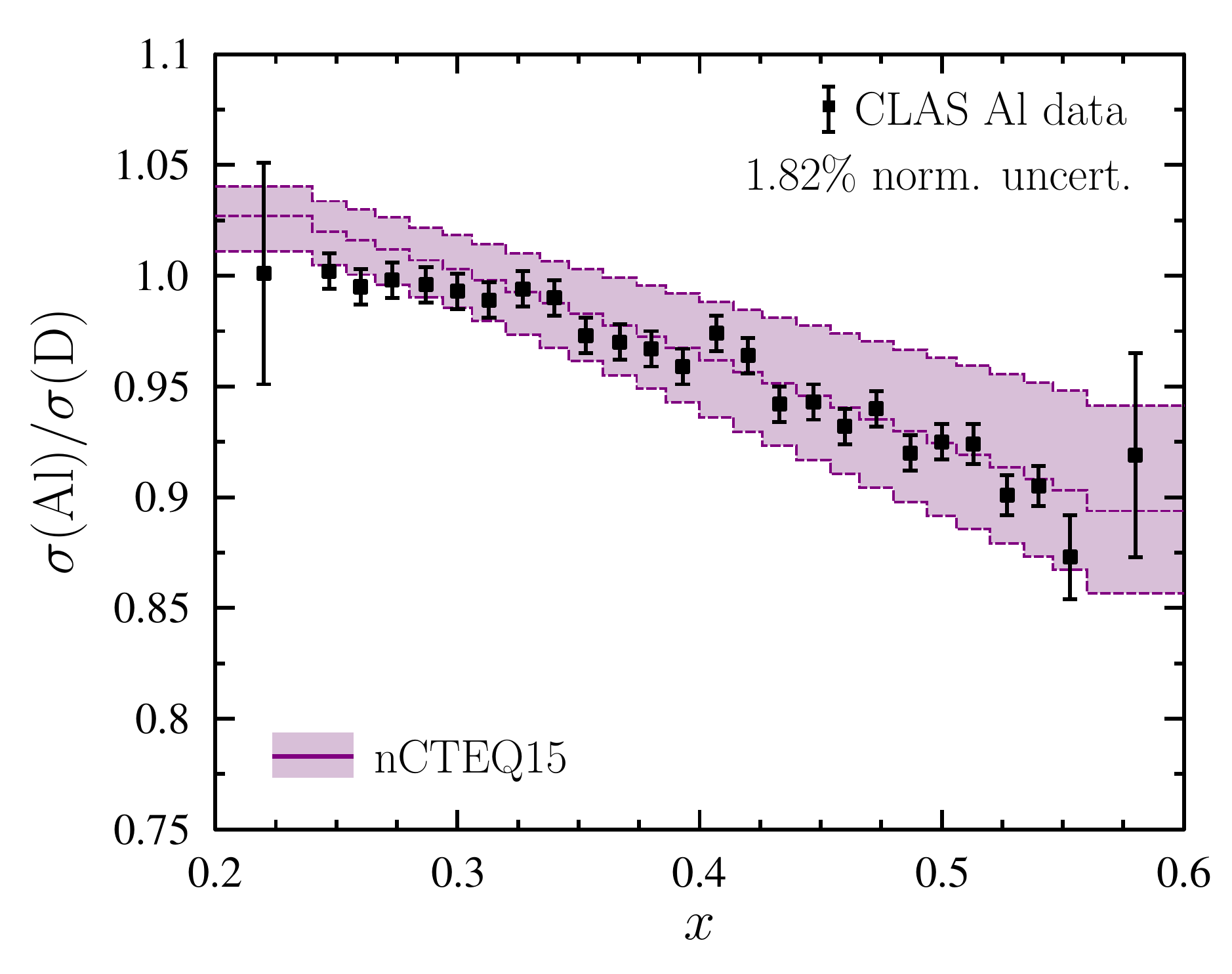}
\includegraphics[width=0.329\linewidth]{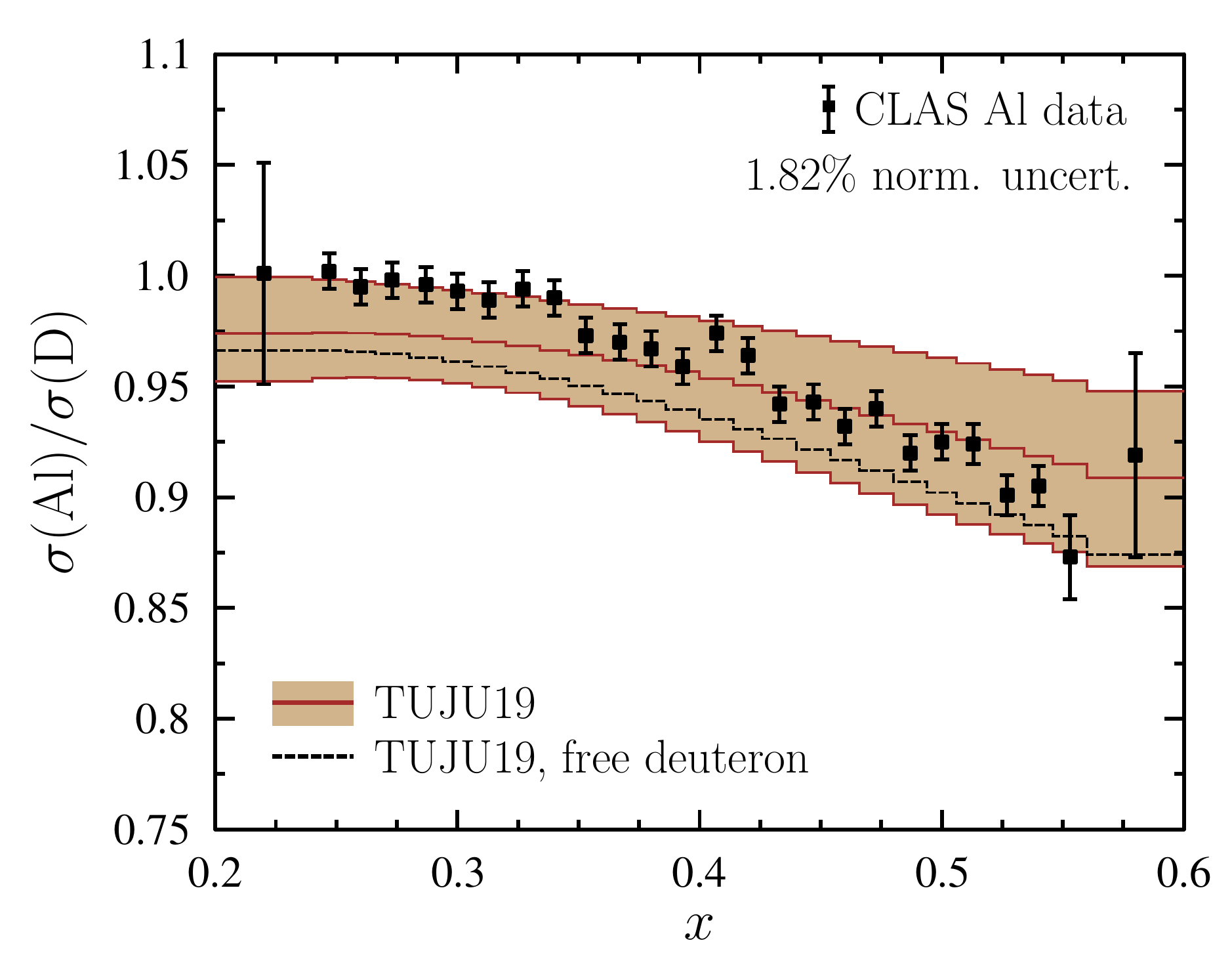}
\includegraphics[width=0.329\linewidth]{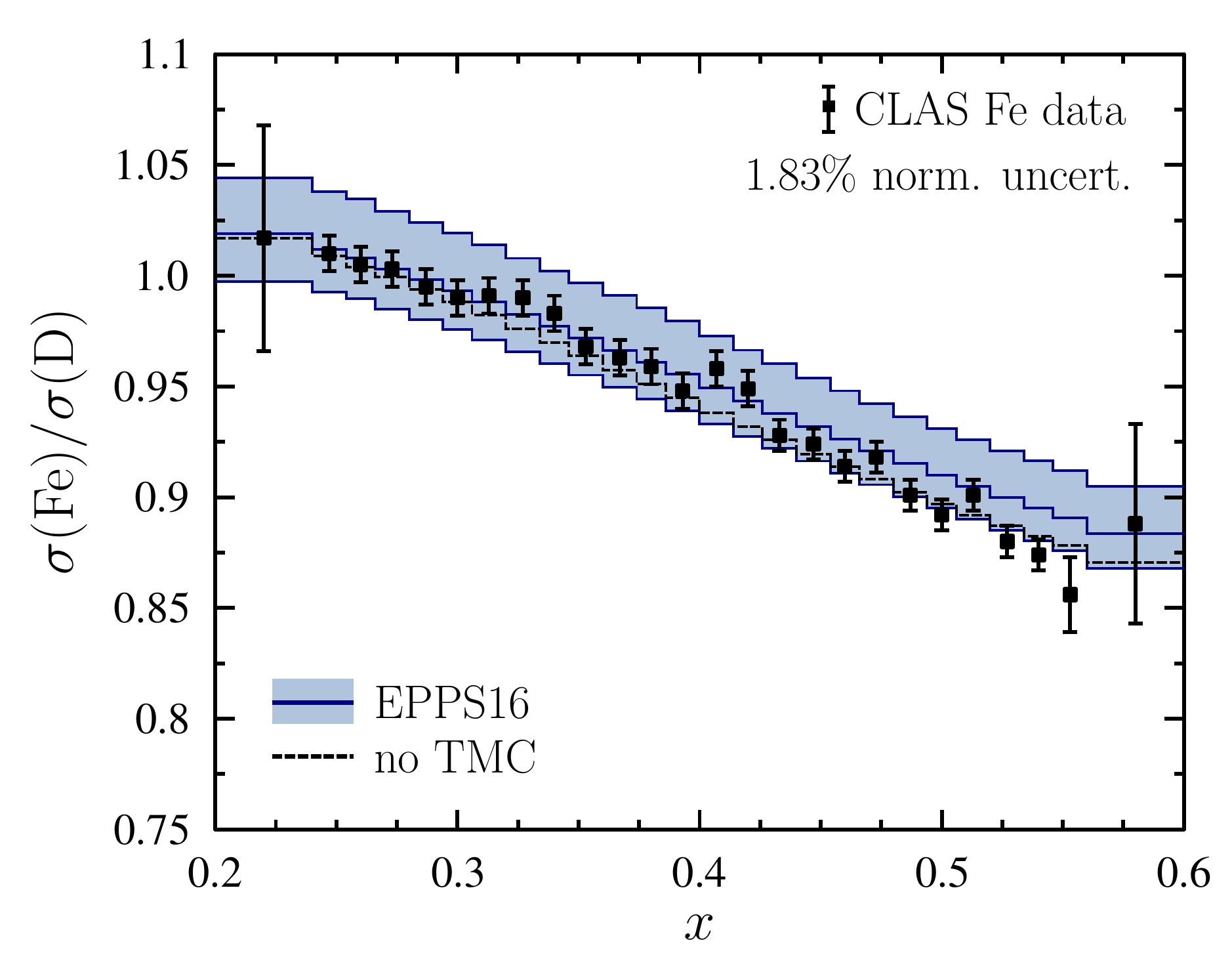}
\includegraphics[width=0.329\linewidth]{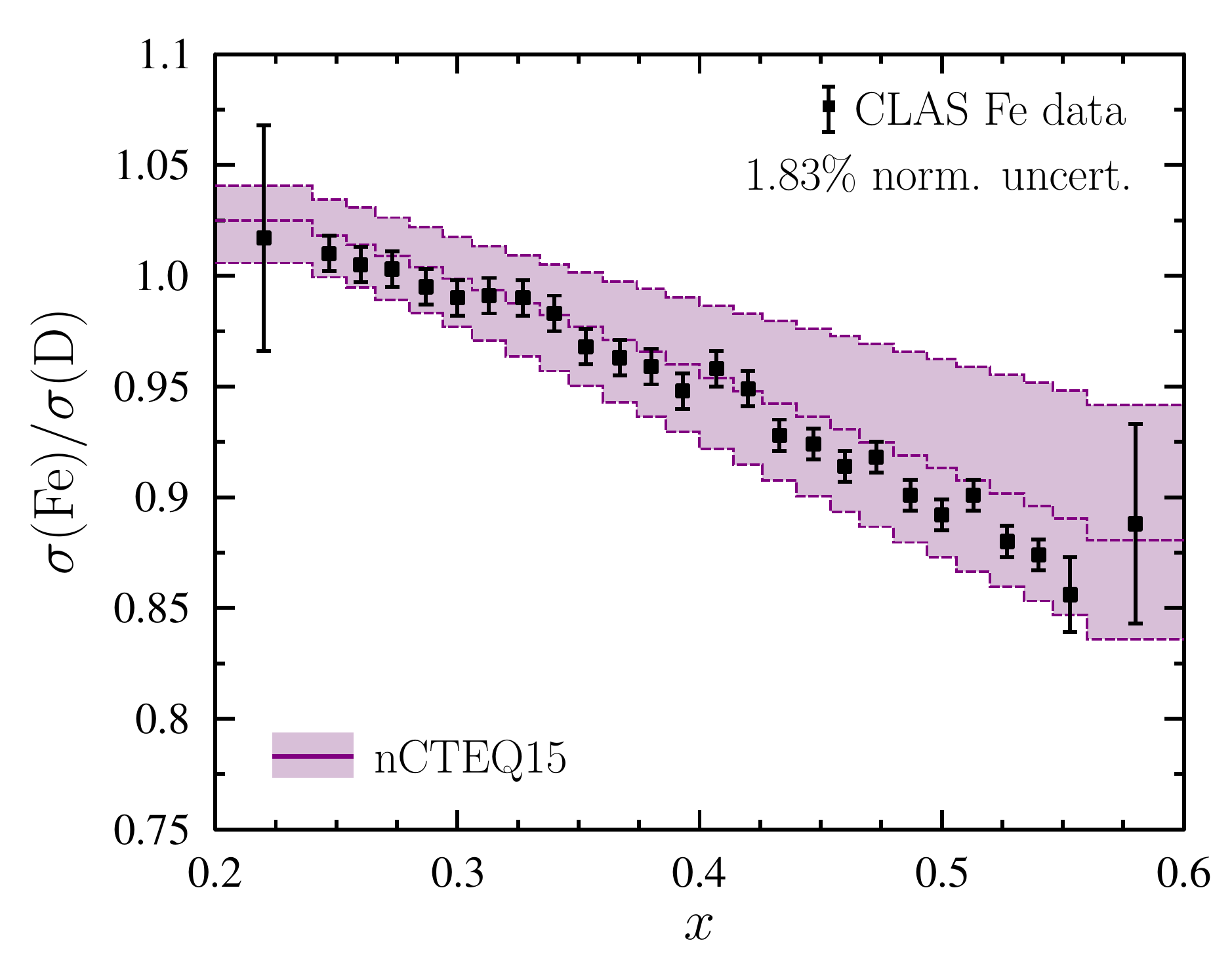}
\includegraphics[width=0.329\linewidth]{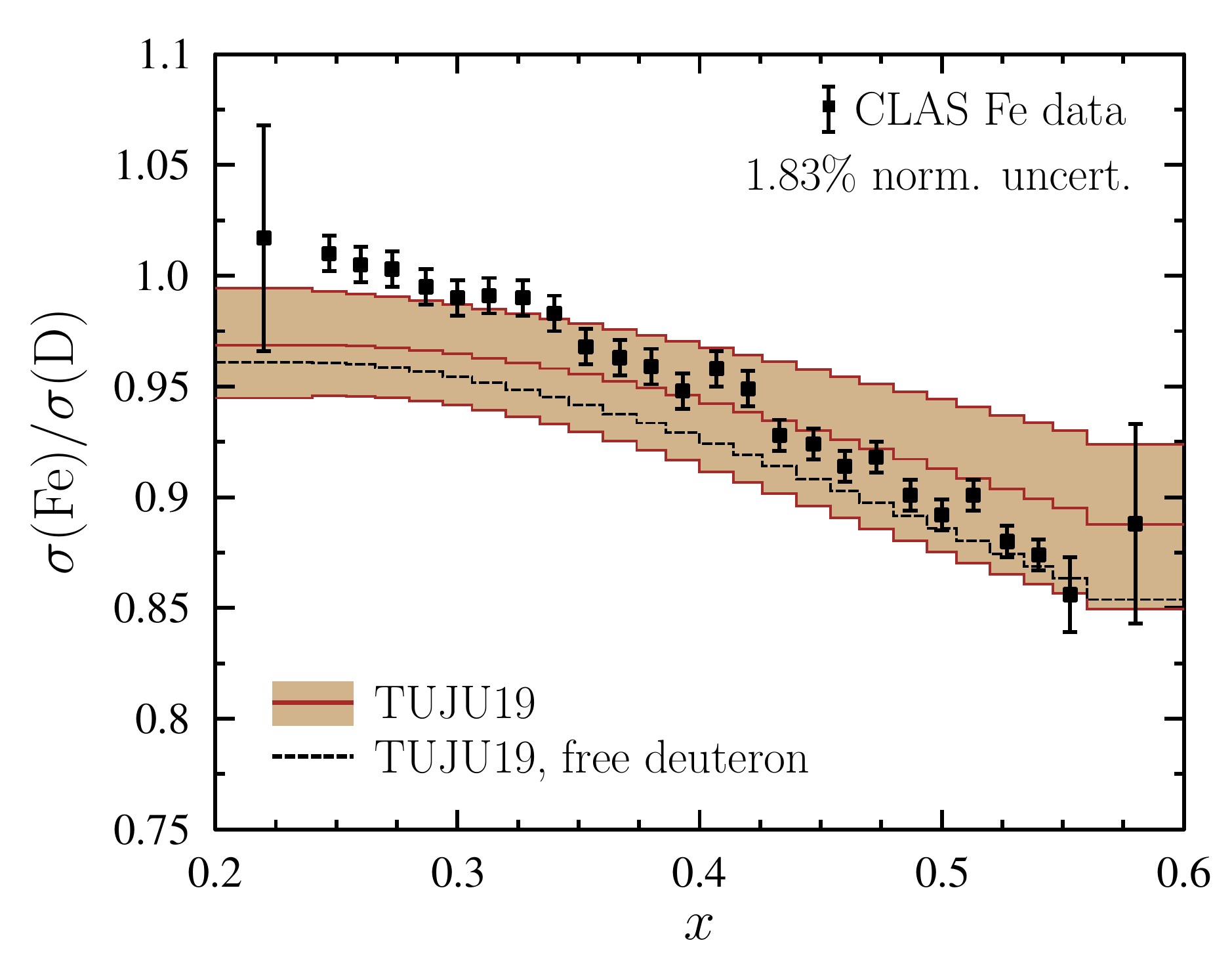}
\includegraphics[width=0.329\linewidth]{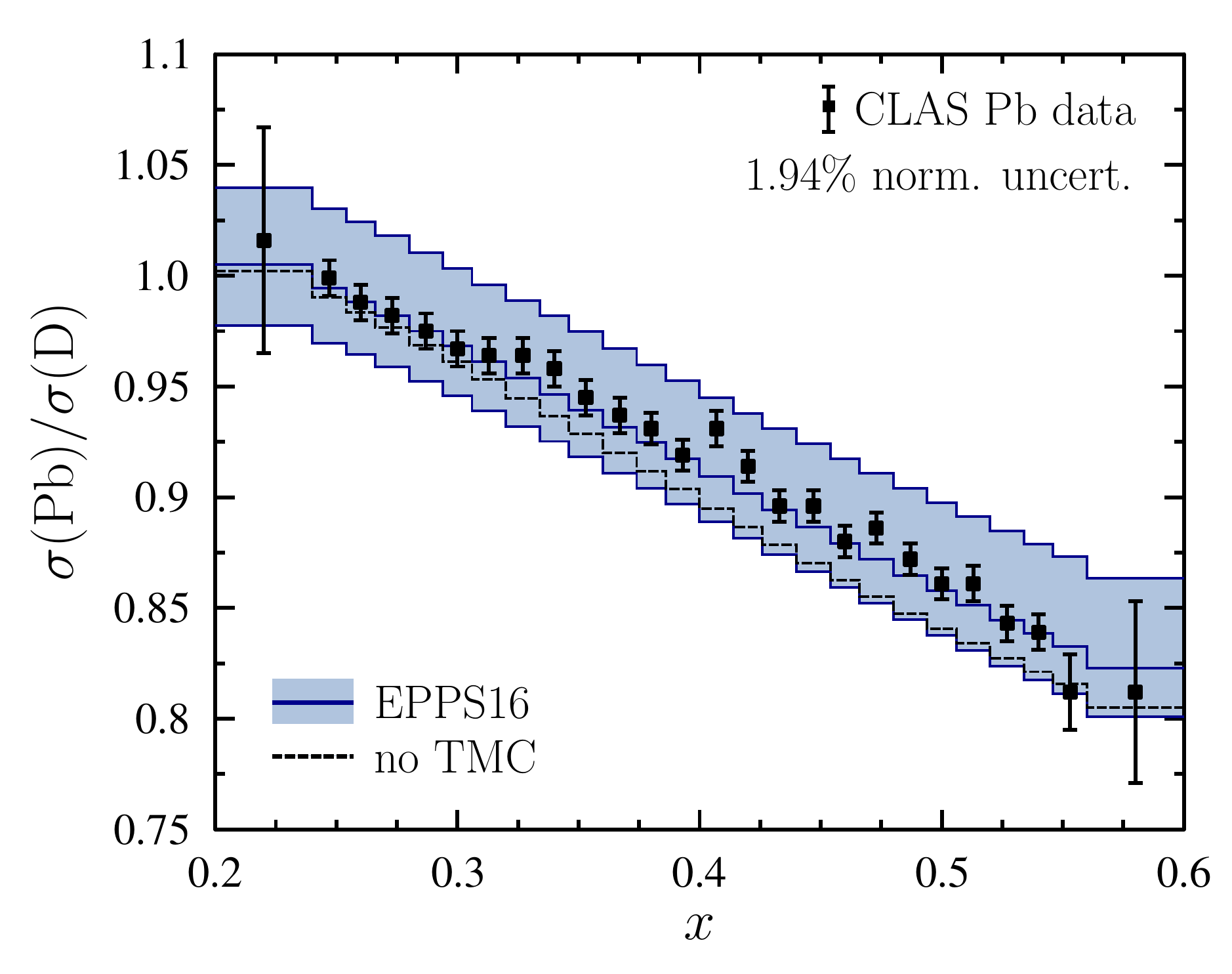}
\includegraphics[width=0.329\linewidth]{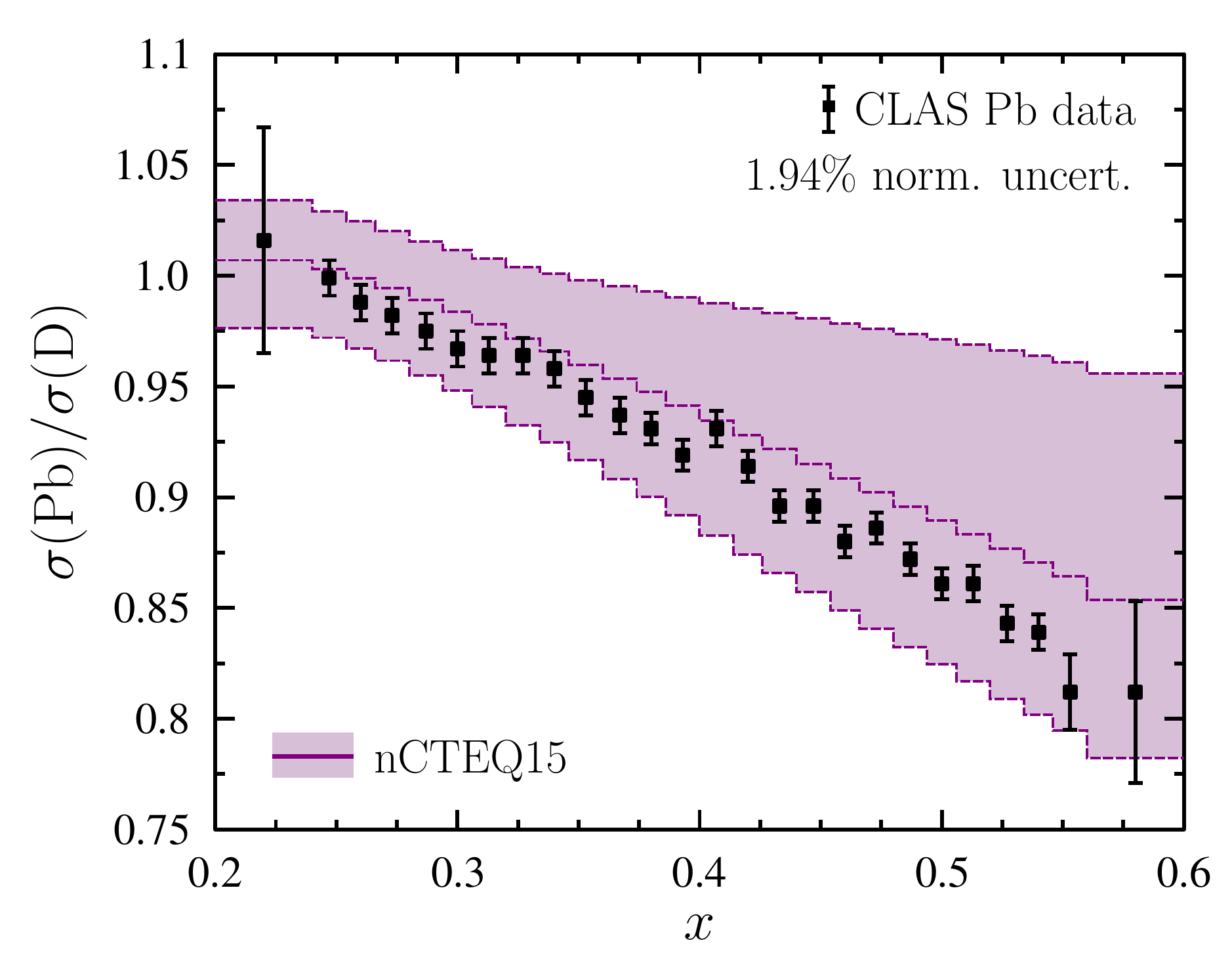}
\includegraphics[width=0.329\linewidth]{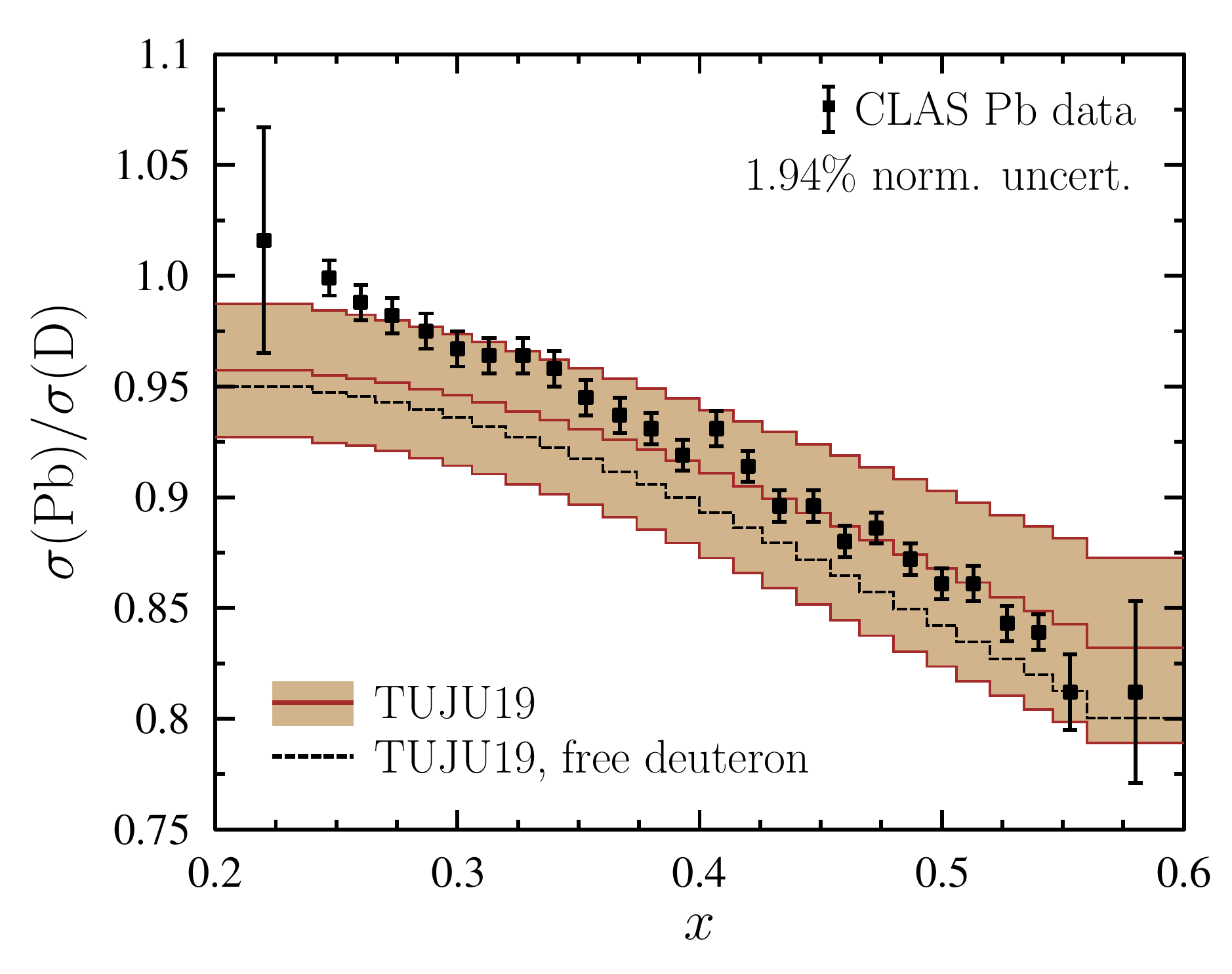}
\caption{The CLAS data compared with the nuclear-PDF predictions. Left panels: EPPS16 with (solid line) and without (dashed line) TMCs. Center panels: nCTEQ15. Right panels: TuJU19 with (solid line) and without (dashed line) nuclear effects in deuteron.}
\label{fig:orig}
\end{figure*} 

The underlying idea is to simulate a global analysis by defining a global $\chi^2$ function
\begin{equation}
\chi^2_{\rm global} \equiv \Delta \chi^2 \sum_k z_k^2 + \chi^2_{\rm new}(\vec z) \,, \label{eq:newchi2}
\end{equation}
where the first term approximates all the data included in a given PDF analysis. The Hessian error sets distributed along published PDFs effectively parametrize the PDFs as a function of the coordinates $z_k$ and can be used to approximate the latter term. The coordinates $z_k$ that minimize $\chi^2_{\rm global}$ then define a new set of PDFs. The first term in Eq.~(\ref{eq:newchi2}) at the minimum is what we call ``PDF penalty" in what follows. Observing a penalty clearly smaller than the tolerance $\Delta \chi^2$ is a sign that the new data can be included in the global analysis without inconsistencies appearing. A penalty larger than $\Delta \chi^2$, in turn, signifies a tension in the PDF fit between the new data and some other data in the original analysis. 
\begin{figure*}[htb!]
\centering
\includegraphics[width=1.0\linewidth]{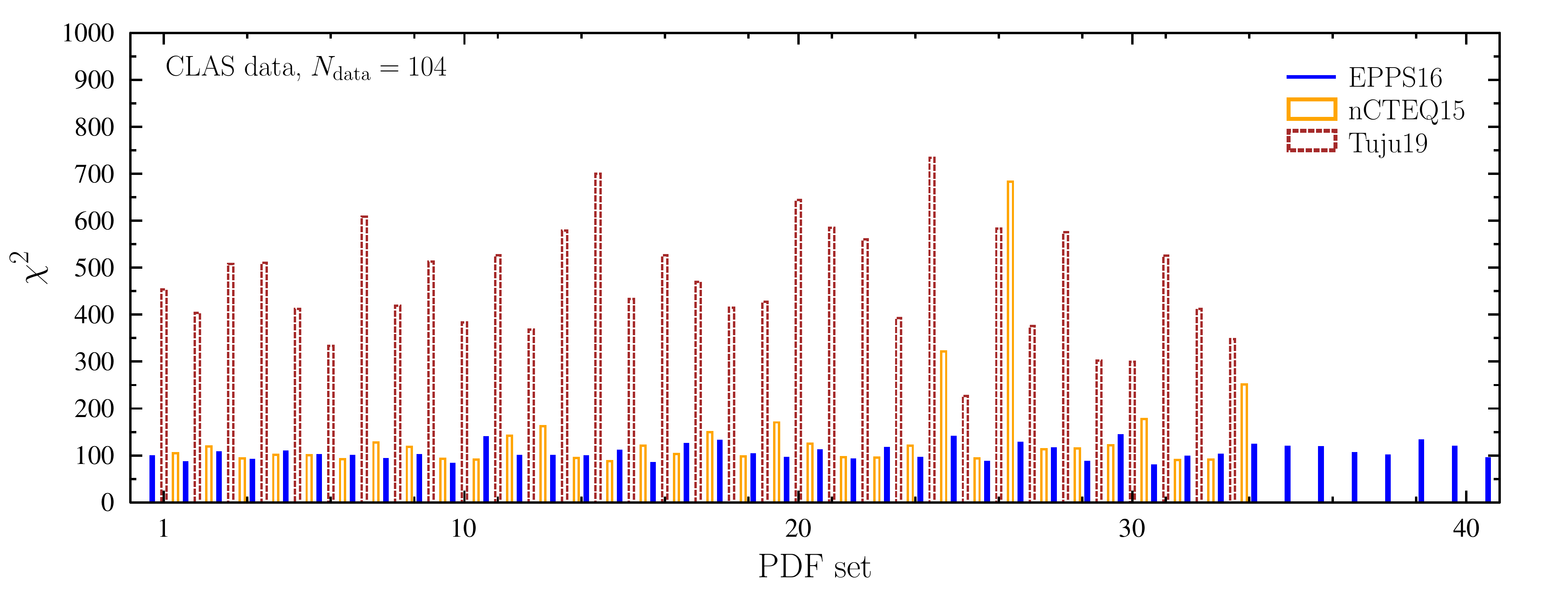}
\caption{The $\chi^2$ values corresponding to the central and error sets of the nuclear PDFs.}
\label{fig:chi2s}
\end{figure*} 
We write our $\chi^2_{\rm new}$ merit-of-figure function as
\begin{align}
\chi^2_{\rm new} & \equiv  \sum_{i} \left[\frac{D_i-T_i-\sum_{k=1}^4 s_k \beta_i^k}{ \delta_i} \right]^2 + \sum_{k=1}^4 s_k^2
\,, \label{eq:chi2new} \\
\beta_i^k & \equiv \delta_{i,k}^{\rm norm.} T_i
\end{align}
where $D_i$ corresponds to central data value and $\delta_i$ is the the uncorrelated point-to-point uncertainty. The relative normalization uncertainties $\delta_{i,k}^{\rm norm.}$ are treated as fully correlated. Note that the systematic shifts $s_k \beta_i^k$ are taken to be proportional to the theory values in order to avoid the D'Agostini bias \cite{Ball:2009qv}. By minimizing the $\chi^2$ with respect to parameters $s_k$ one finds the ``optimum shifts" $s_k^{\rm min} \beta_i^k$ that correspond to a given set of theory predictions $T_i$.


\section{Results}\label{section:results}

The first thing done was to compare the data with the expected predictions from nuclear PDFs. We present the results for all the three nuclear-PDF sets and all four nuclei in Fig.~\ref{fig:orig}. In the case of EPPS16 we also compare to a calculation without TMCs and in the case of TuJu19 to a calculation which assumes no nuclear effects in deuteron. With EPPS16 and nCTEQ15 the agreement is visually quite good: most data lie within the uncertainty bands and the downward slopes are well reproduced. The nCTEQ15 error bands are generally larger than those of EPPS16 -- particularly for Pb -- and can be explained by the larger uncertainties in the up-valence distributions as was seen in Fig.~\ref{fig:valenceall}. This indicates that these data should be able to set significant new constraints especially for nCTEQ15. While the data are also broadly reproduced by the TuJu19 PDFs, the predicted EMC slope appears to be somewhat too flat systematically for all the four nuclei and the predictions tend to underestimate the data around $0.2<x<0.35$. While perhaps a bit unexpected, the systematic difference between the EPPS16/nCTEQ15 and TuJu19 predictions is consistent e.g. with Fig.~10 of the original TuJu19 paper \cite{Walt:2019slu}, where the fit can be seen to somewhat underestimate the NMC data for C/D and Ca/D ratios at $x \gtrsim 0.1$. The EPPS16 values for these same data are somewhat higher, as can be seen from Fig.~13 of the original EPPS16 paper \cite{Eskola:2016oht}, and better agree with the data. Thus, the differences we observe here seem to be consistent. 

\begin{figure*}[htb!]
\centering
\includegraphics[width=0.329\linewidth]{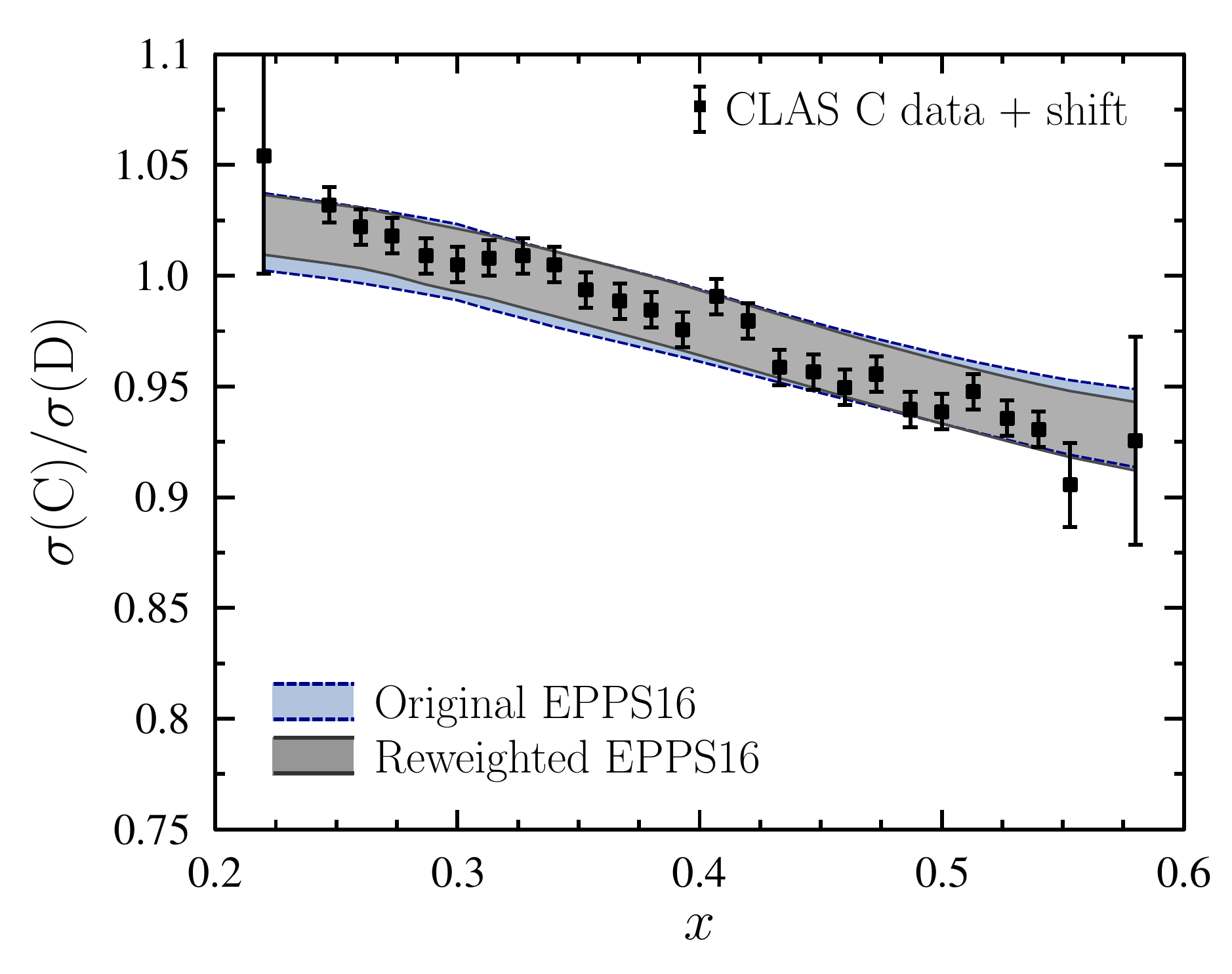}
\includegraphics[width=0.329\linewidth]{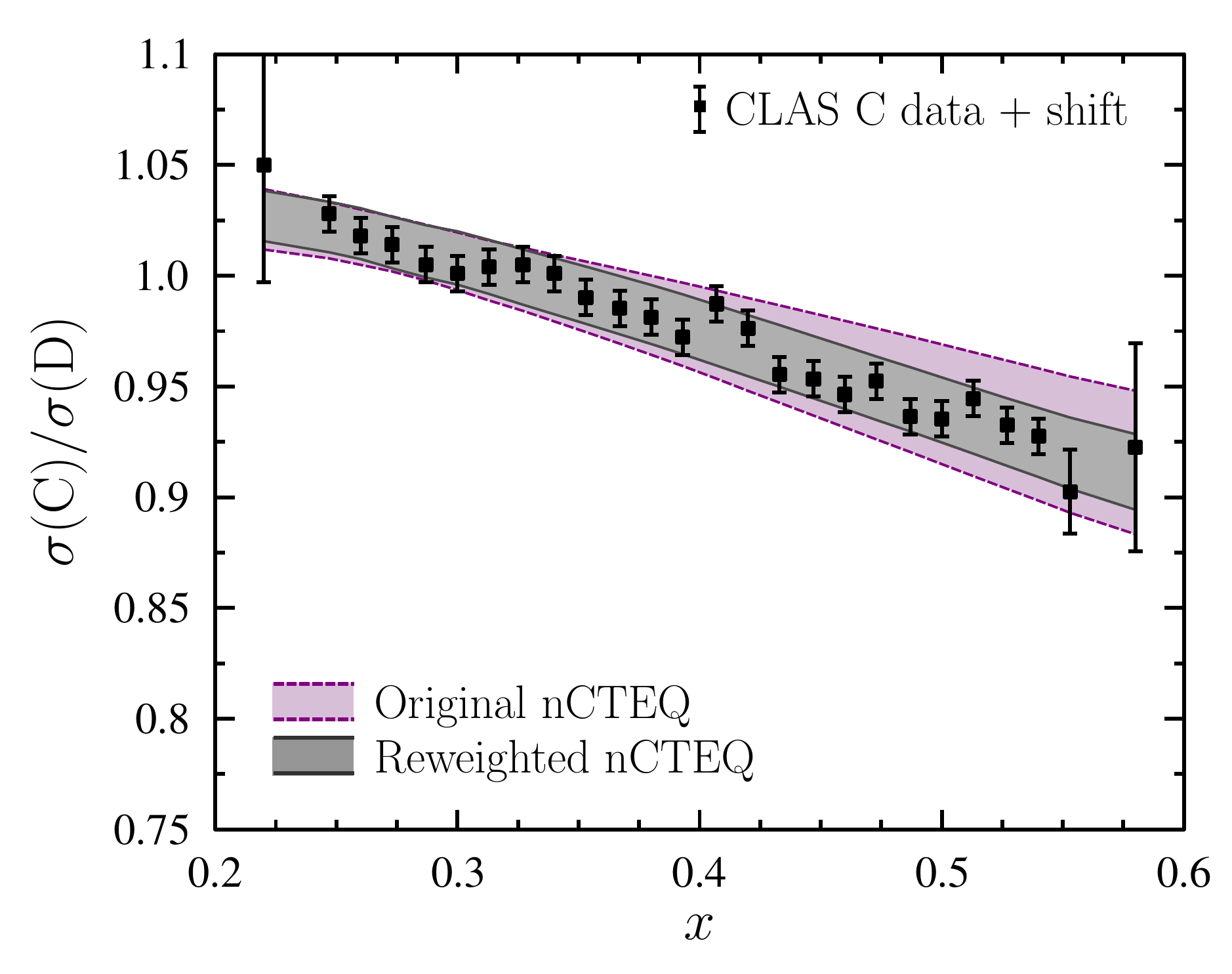}
\includegraphics[width=0.329\linewidth]{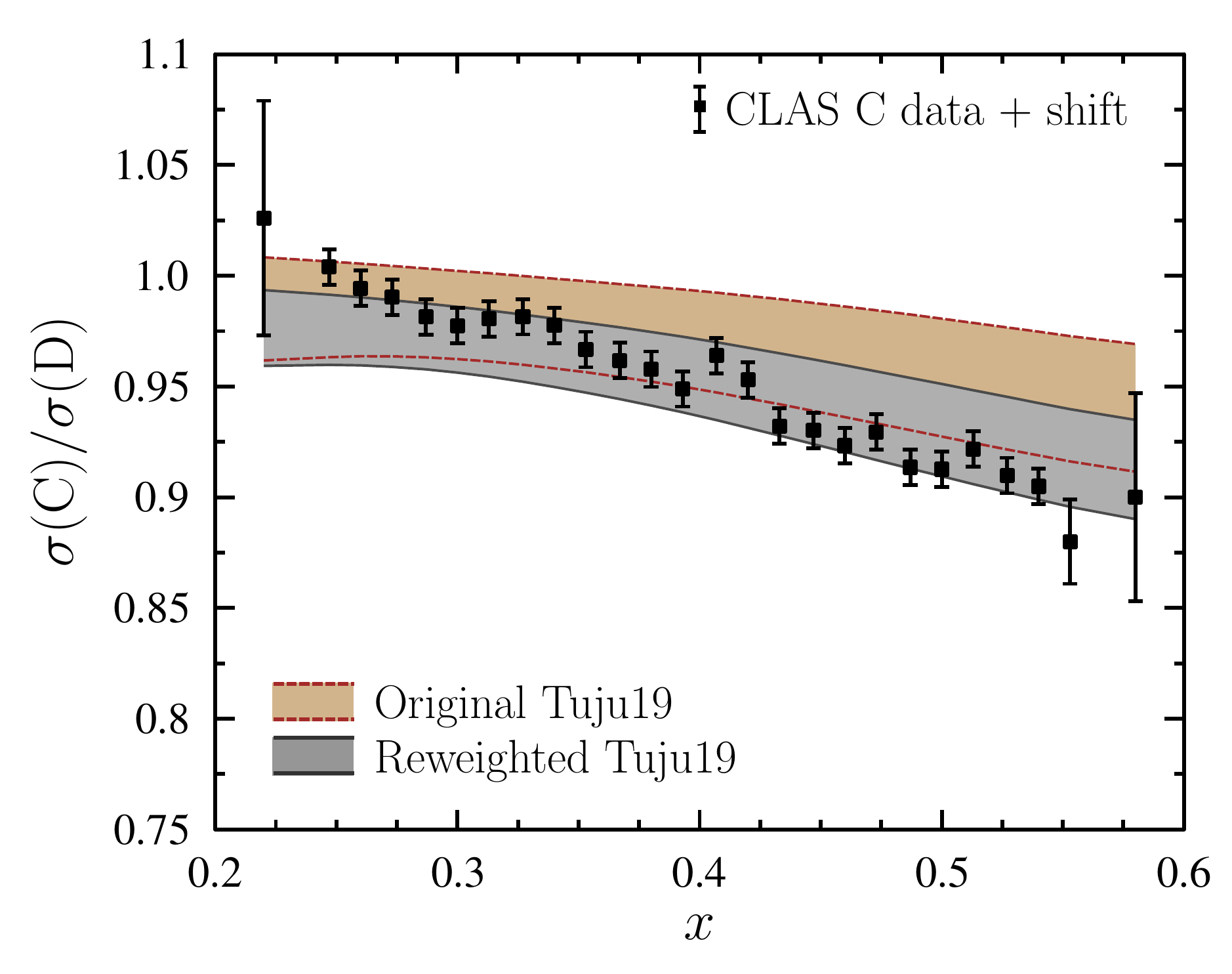}
\includegraphics[width=0.329\linewidth]{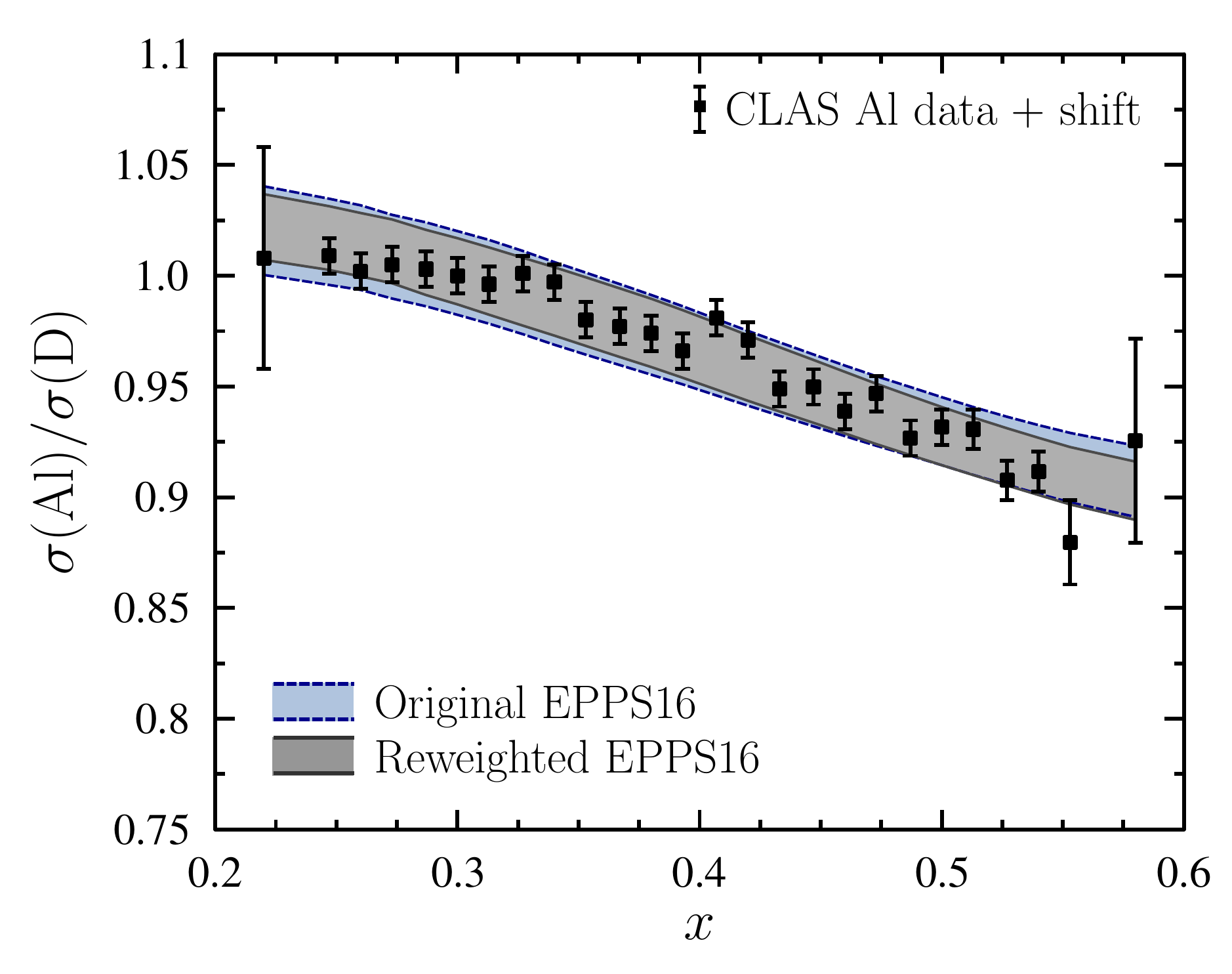}
\includegraphics[width=0.329\linewidth]{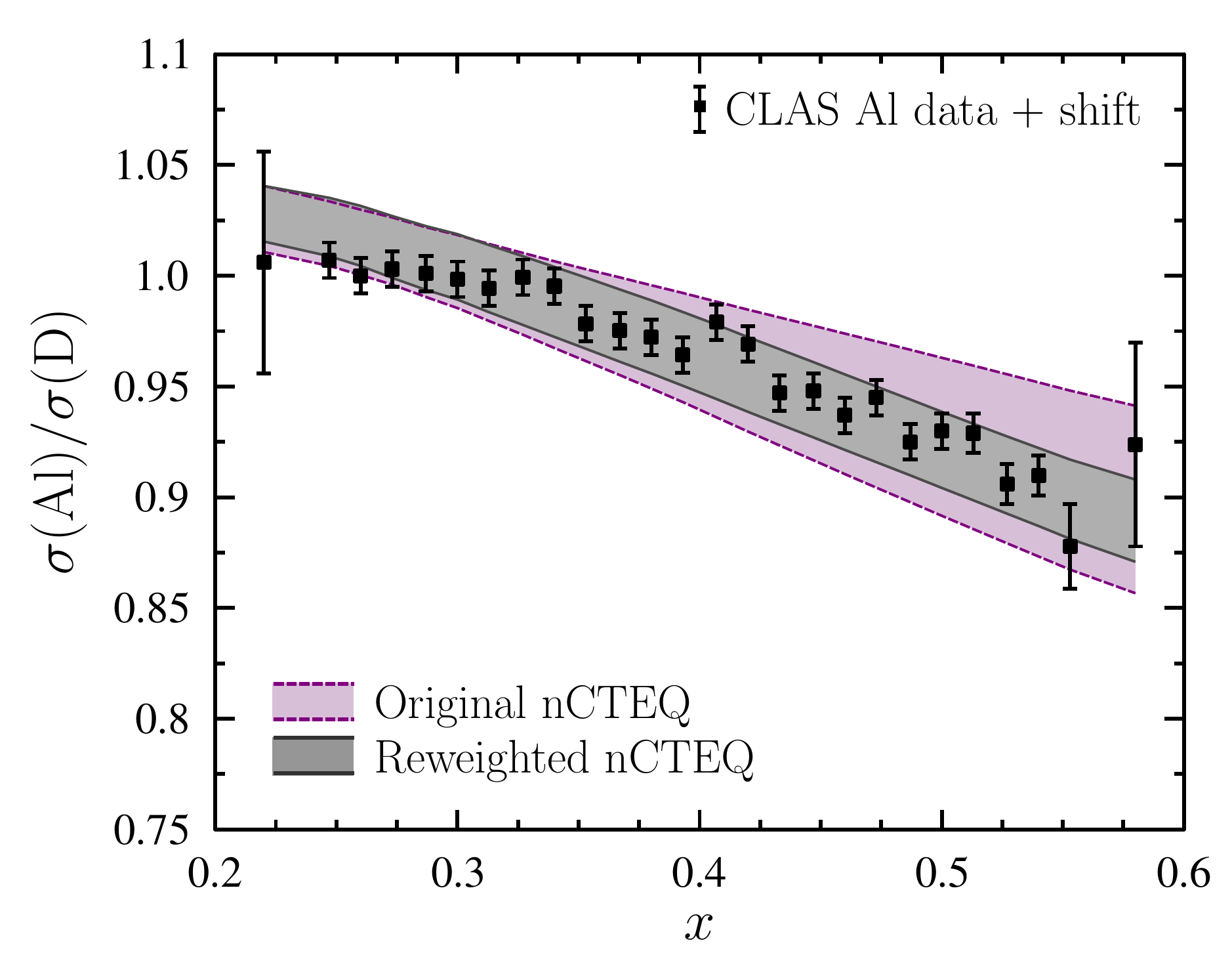}
\includegraphics[width=0.329\linewidth]{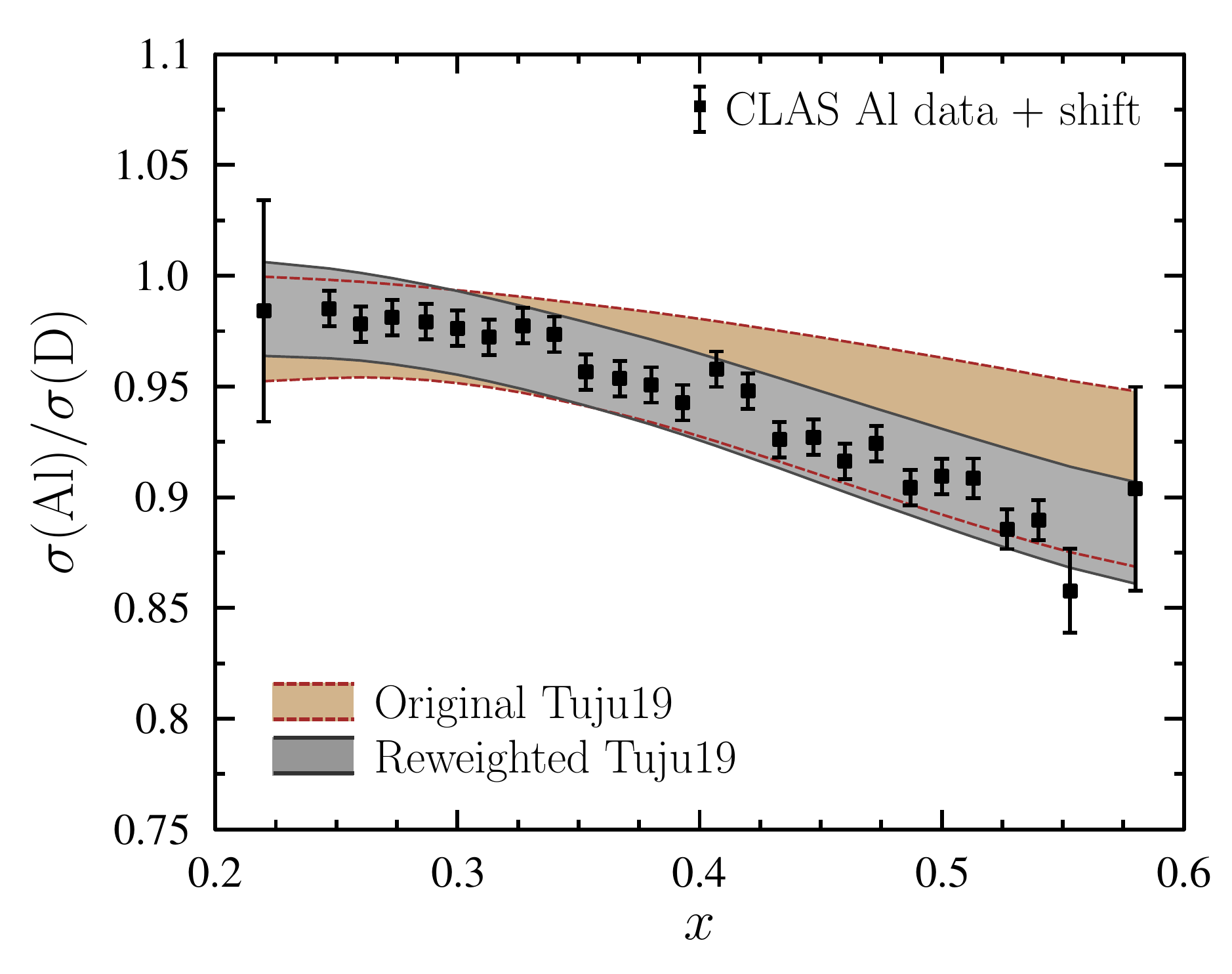}
\includegraphics[width=0.329\linewidth]{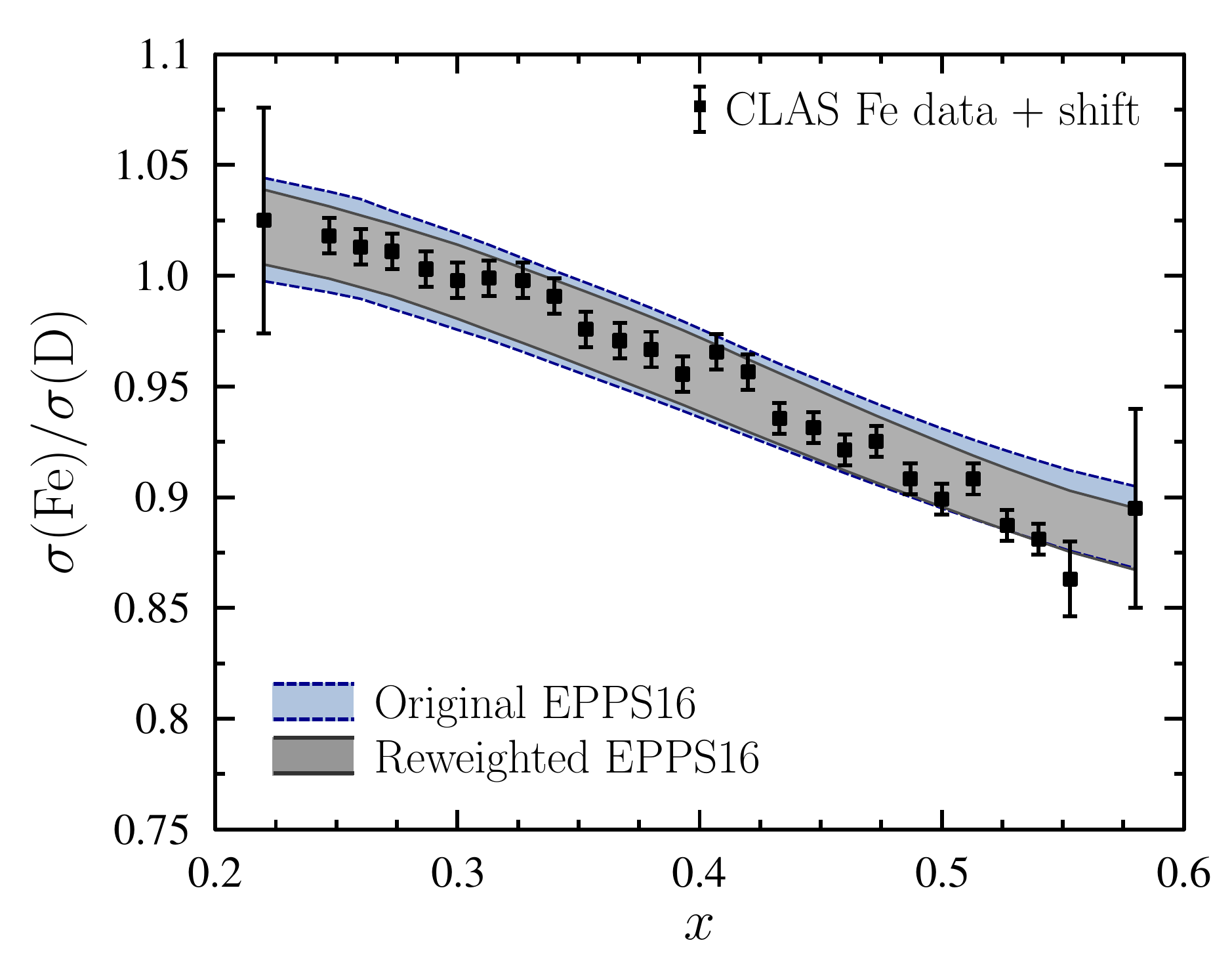}
\includegraphics[width=0.329\linewidth]{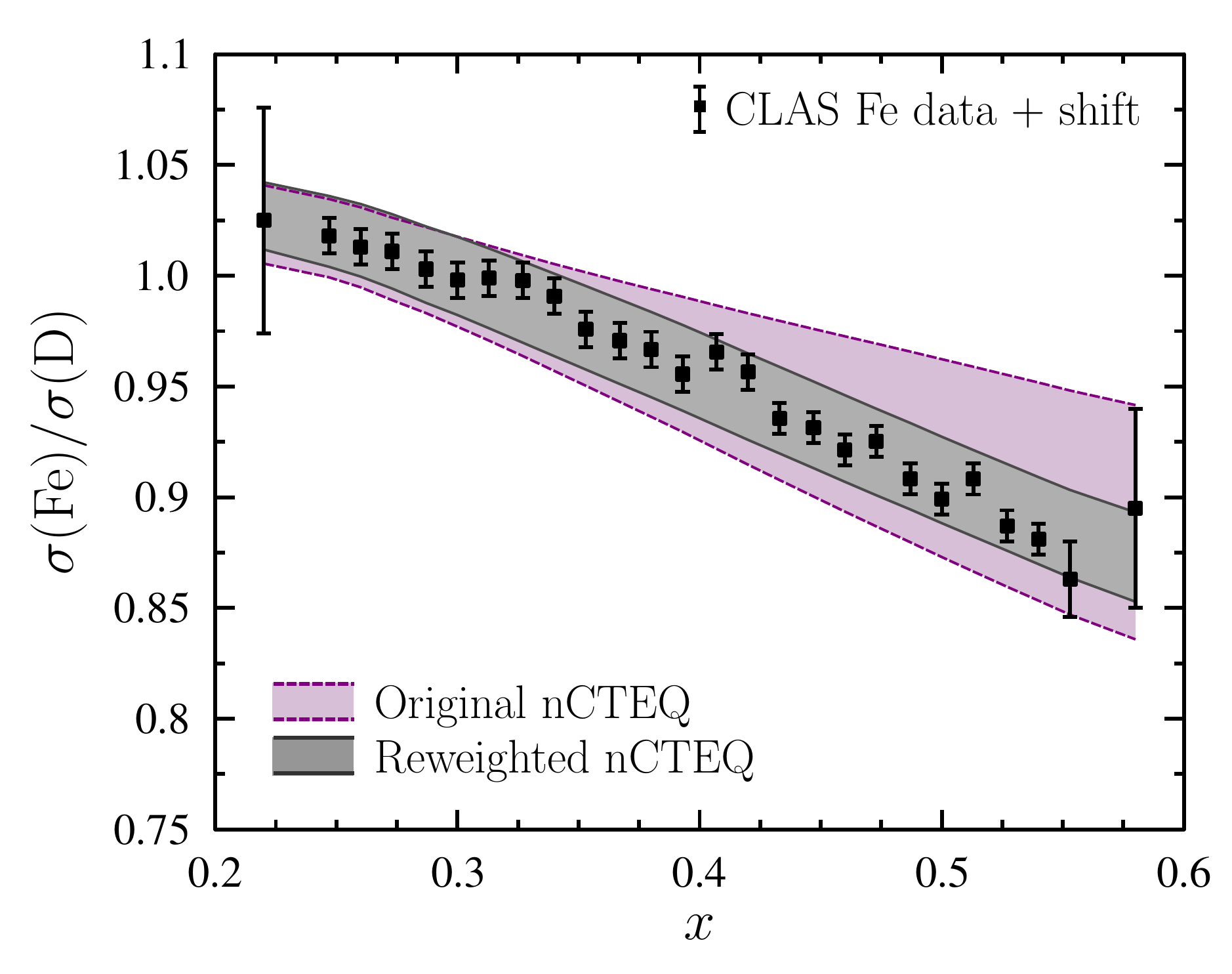}
\includegraphics[width=0.329\linewidth]{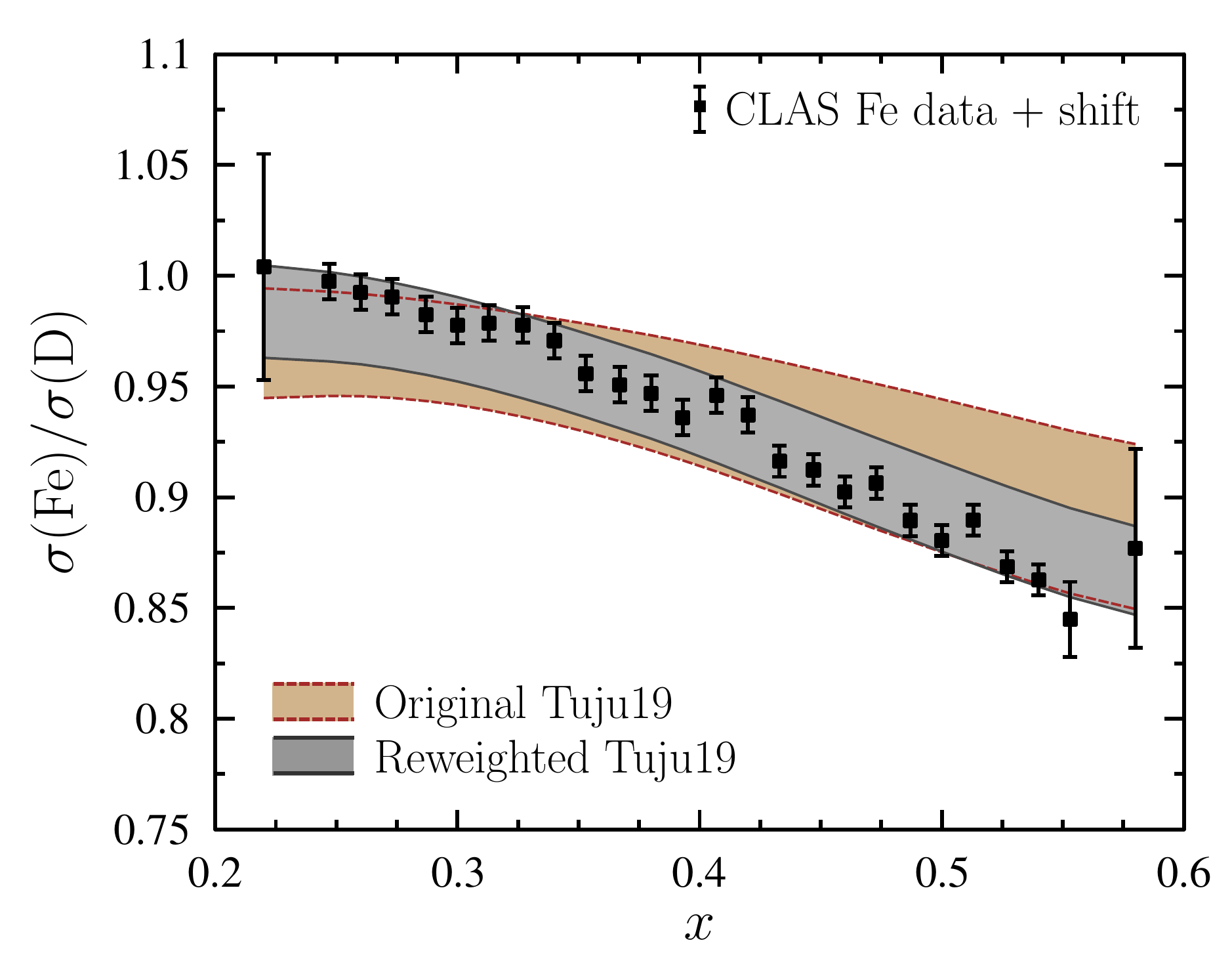}
\includegraphics[width=0.329\linewidth]{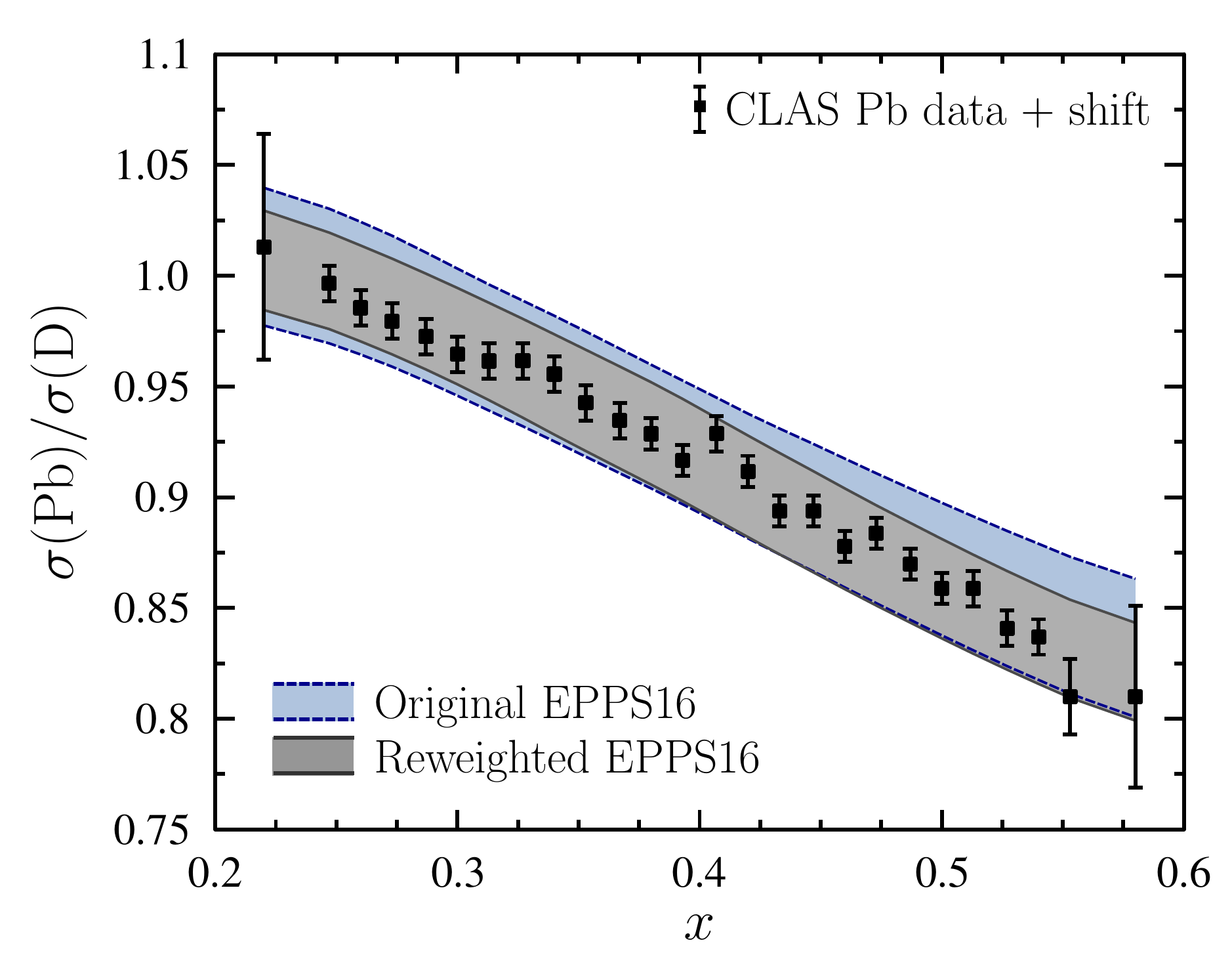}
\includegraphics[width=0.329\linewidth]{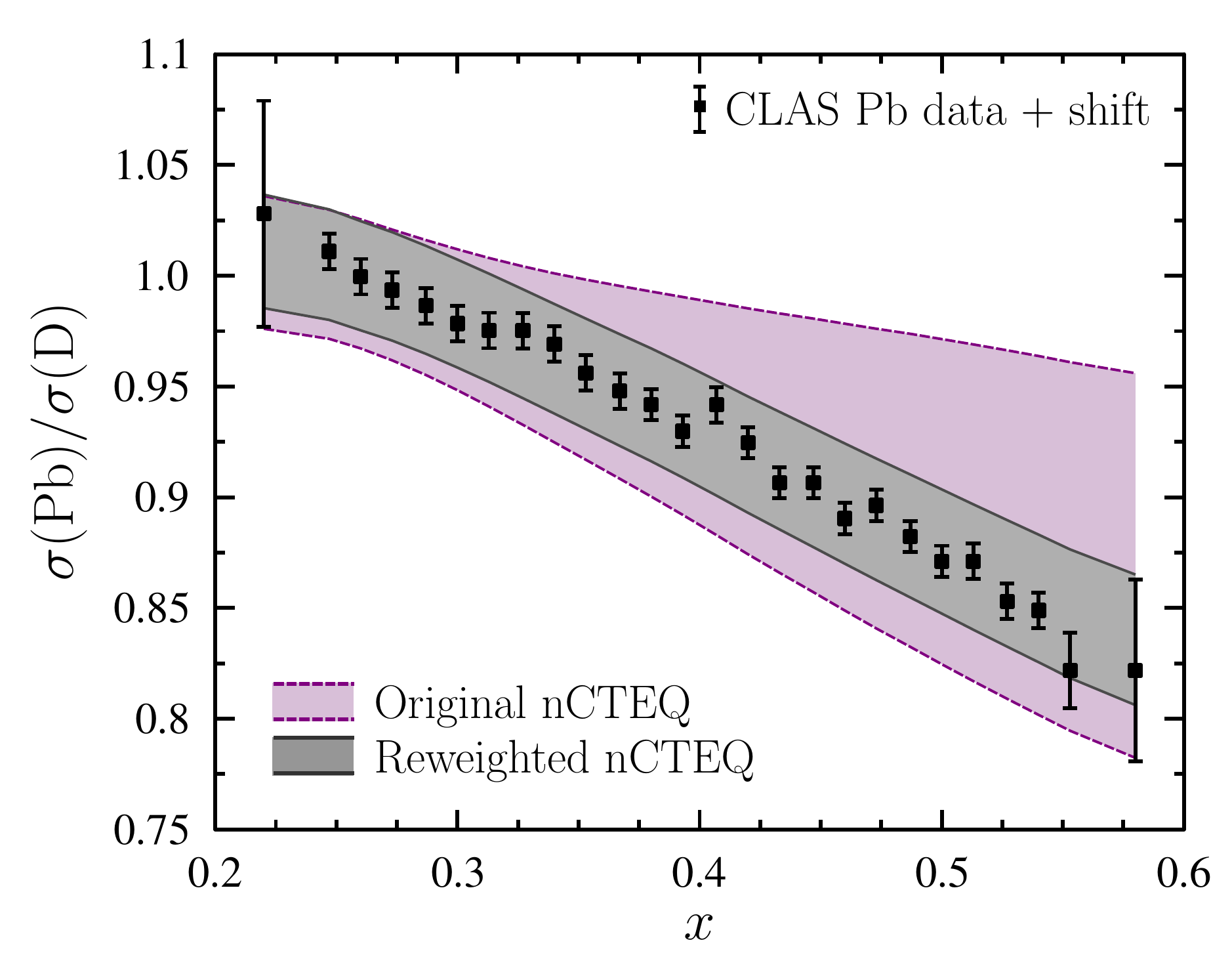}
\includegraphics[width=0.329\linewidth]{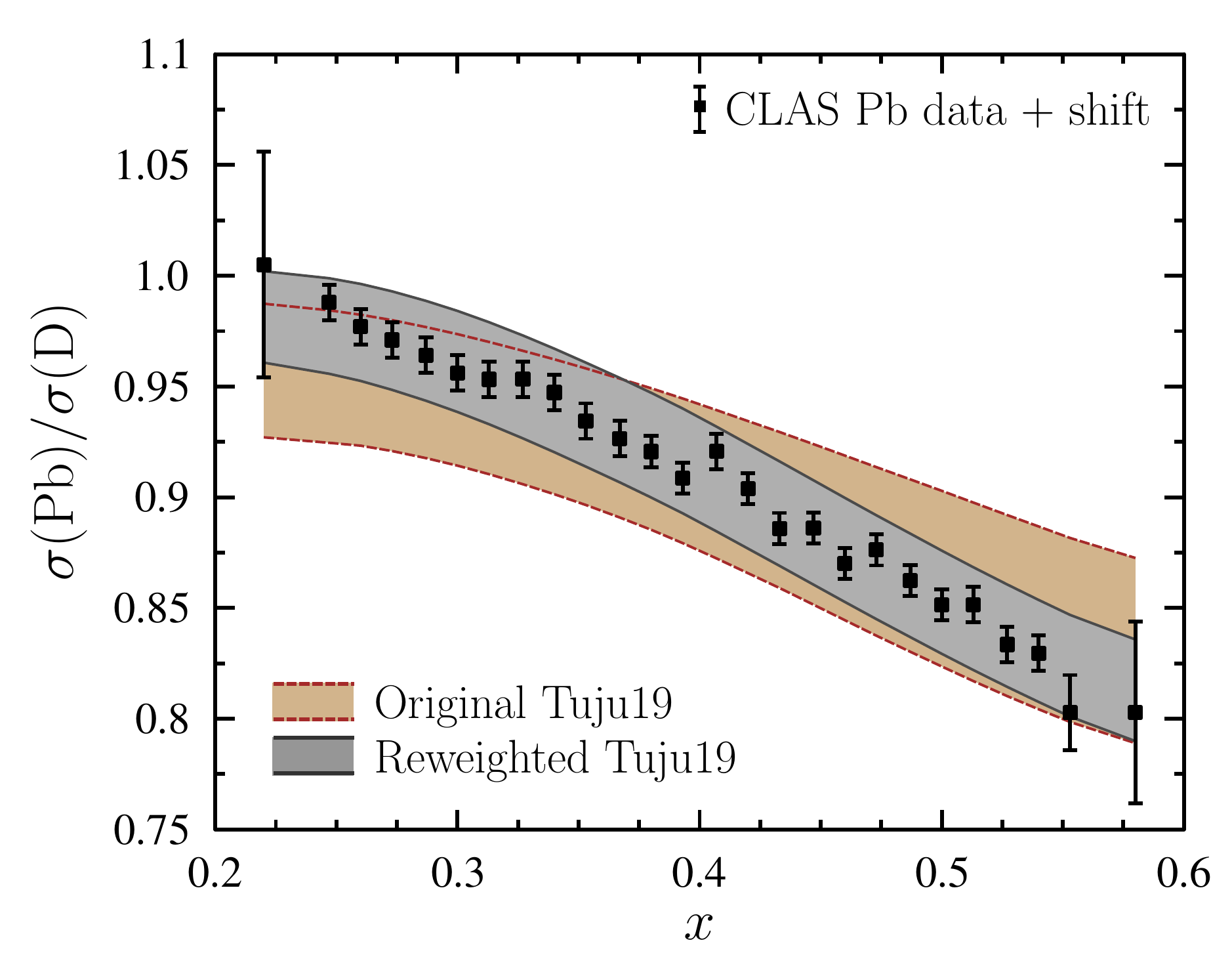}
\caption{The CLAS data compared with the re-weighted nuclear-PDF predictions. The optimal shifts that minimize Eq.~(\ref{eq:chi2new}) with the central re-weighted predictons, have been applied to the data points.}
\label{fig:reweighteddata}
\end{figure*} 

From the EPPS16 panels of Fig.~\ref{fig:orig} we see that the effect of TMCs becomes relevant at $x \gtrsim 0.3$ and the TMCs evidently provoke and upward shift in the predictions. Since the Nachtmann variable $\xi$ is always smaller than Bjorken $x$, $\xi < x$, by turning on the TMCs one effectively probes the nPDFs at bit lower momentum fraction. As can be seen from Fig.~\ref{fig:valenceall}, the nuclear effects in PDFs are monotonic in the EMC region so by shifting to a smaller momentum fraction by turning on TMCs the cross-section ratios increase a bit. It appears that a slightly better agreement with the data is obtained with TMCs -- we will later on see to what extent this is significant. In any case, the effect of TMCs competes with the uncorrelated data uncertainties so it might become relevant, then, to consider TMCs in future fits of nuclear PDFs.

Out of the three nuclear-PDF fits considered here, the TuJu19 analysis is the only one to consider nuclear effects for deuteron. This was done by extending the parametrization of the $A$ dependence down to $A=2$ and utilizing deuteron structure-function data as a constraint. The effect of nuclear corrections to deuteron PDFs are indicated in the TuJu19 panels of Fig.~\ref{fig:orig}. The corrections are the largest at the highest values of $x$, amounting to $\sim 4\%$ at the most. This appears to be in line with e.g. the phenomenological study of Ref.~\cite{Martin:2012da}. The estimated effects of deuteron corrections exceed the uncorrelated data uncertainties at $x \gtrsim 0.35$. The EPPS16 and nCTEQ15 analyses do not consider nuclear effects for deuteron basically because the smooth, power-law type parametrization of the $A$ dependence may not be completely reliable for very small nuclei, but some discontinuities could be expected at small $A$. For example, the HKN07 analysis \cite{Hirai:2007sx} introduced an extra overall parameter to suppress the otherwise somewhat too strong modifications of the deuteron PDFs. In fact, a possible explanation why TuJu19 fails to reproduce the CLAS data is that the parametrization of the $A$ dependence is too simple to reliably cover all considered nuclei.

\begin{figure*}[htb!]
\centering
\includegraphics[width=0.329\linewidth]{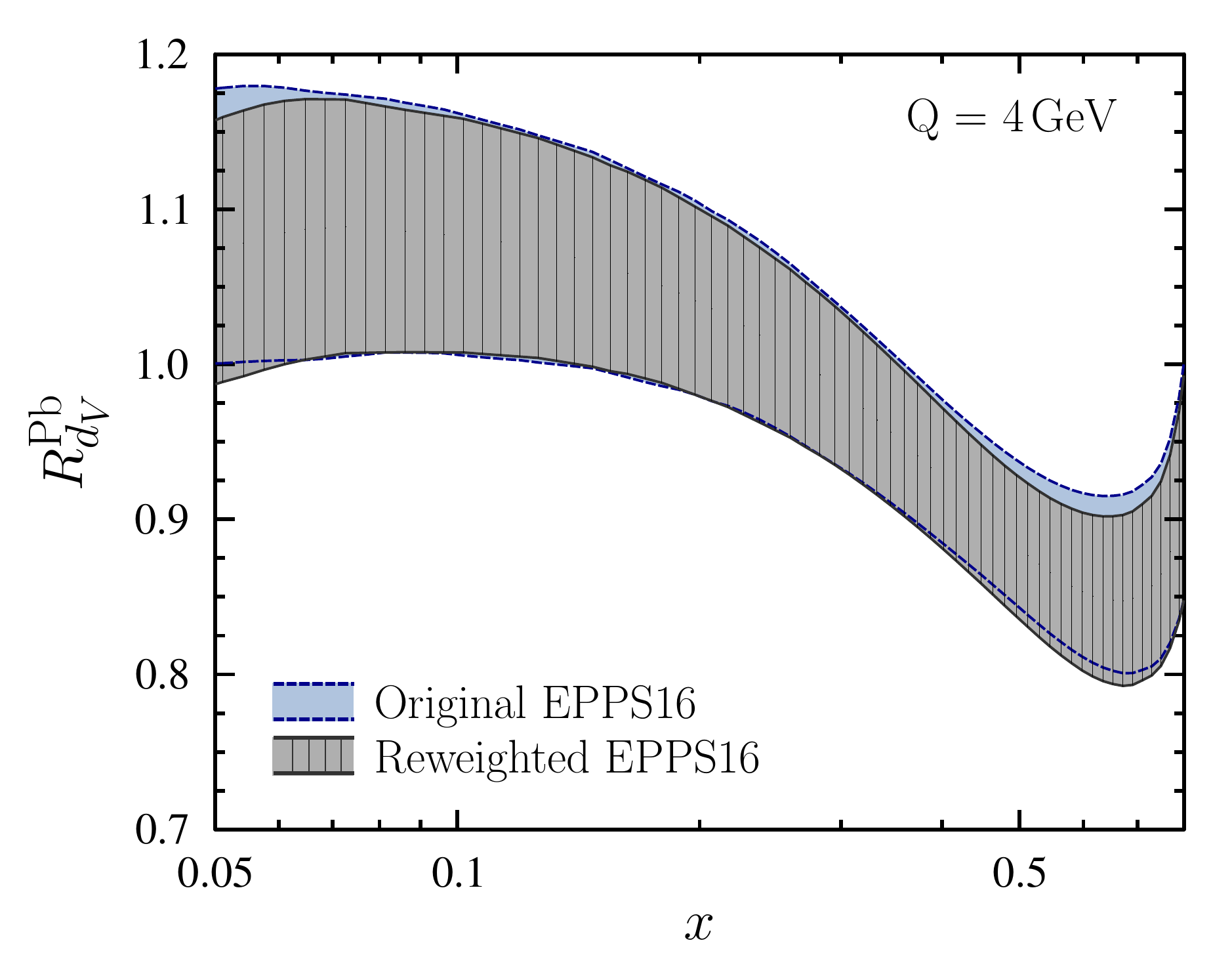}
\includegraphics[width=0.329\linewidth]{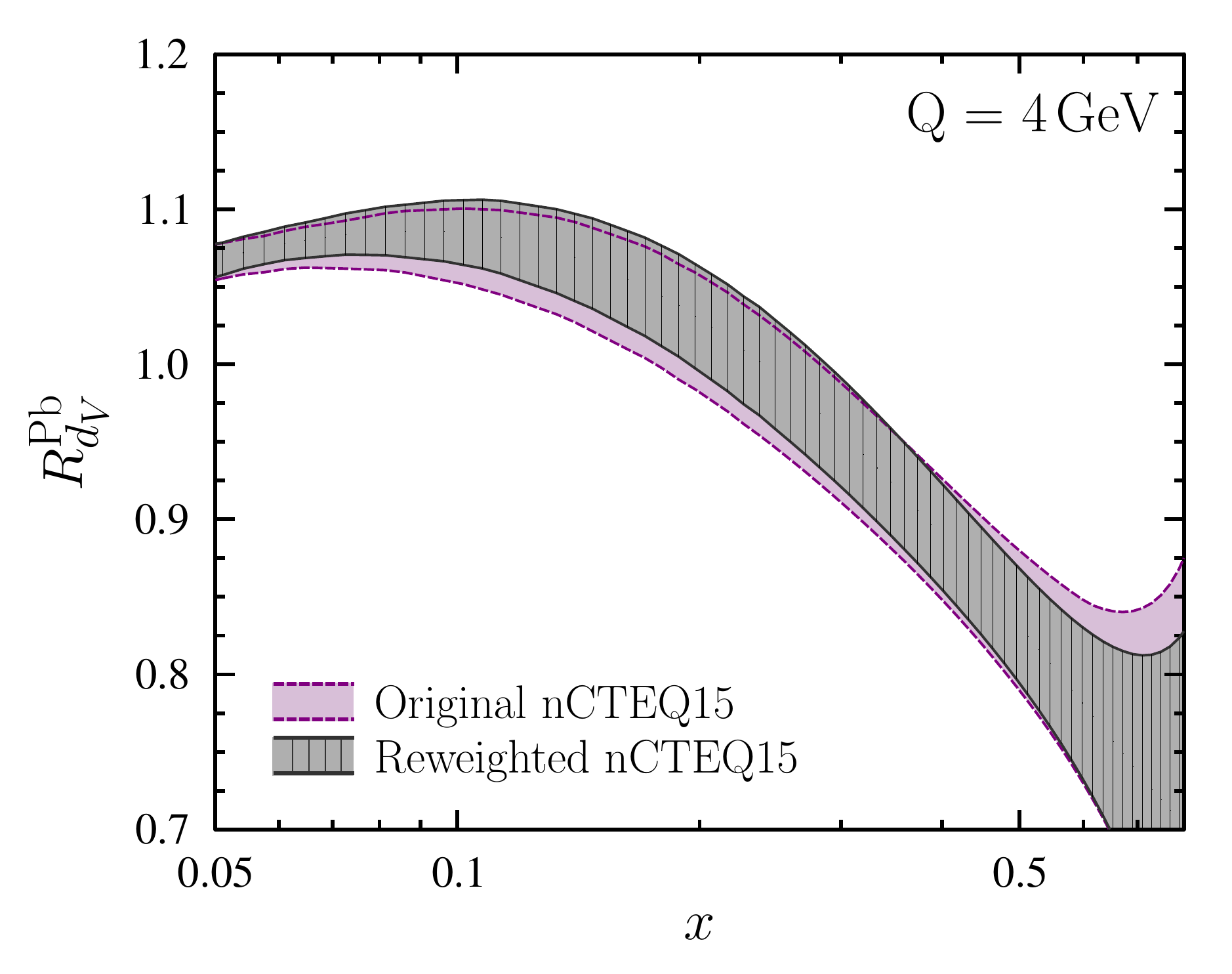}
\includegraphics[width=0.329\linewidth]{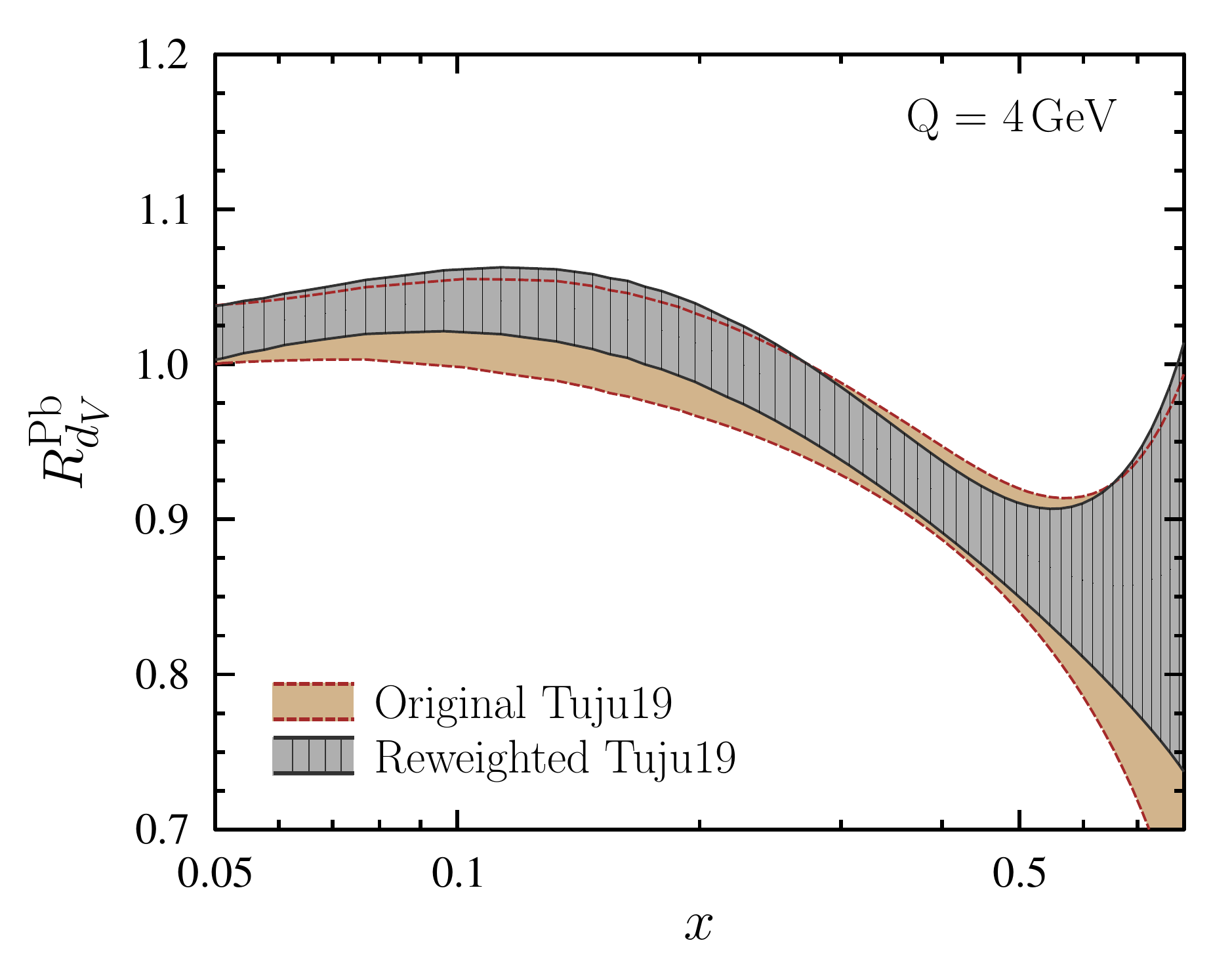}
\includegraphics[width=0.329\linewidth]{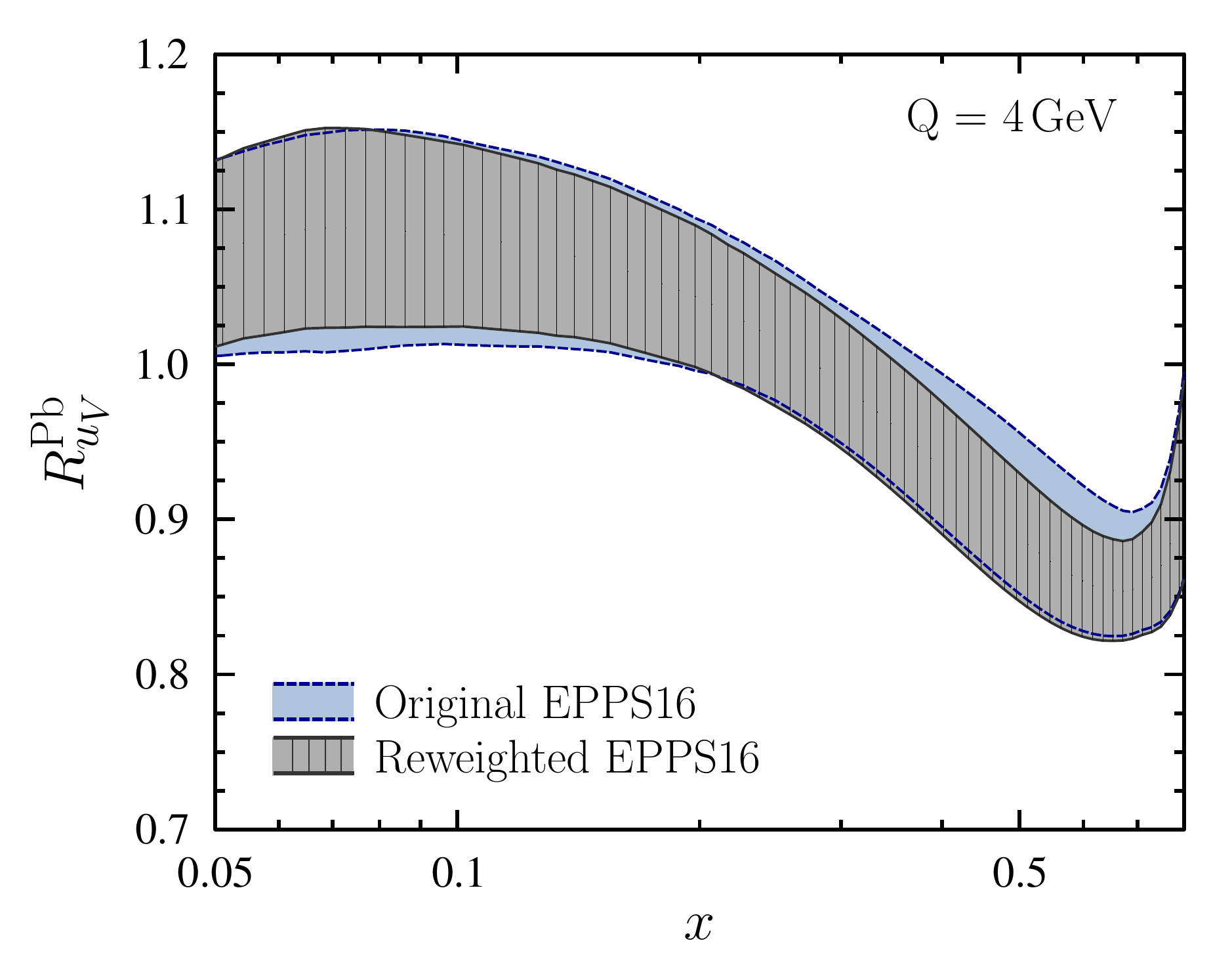}
\includegraphics[width=0.329\linewidth]{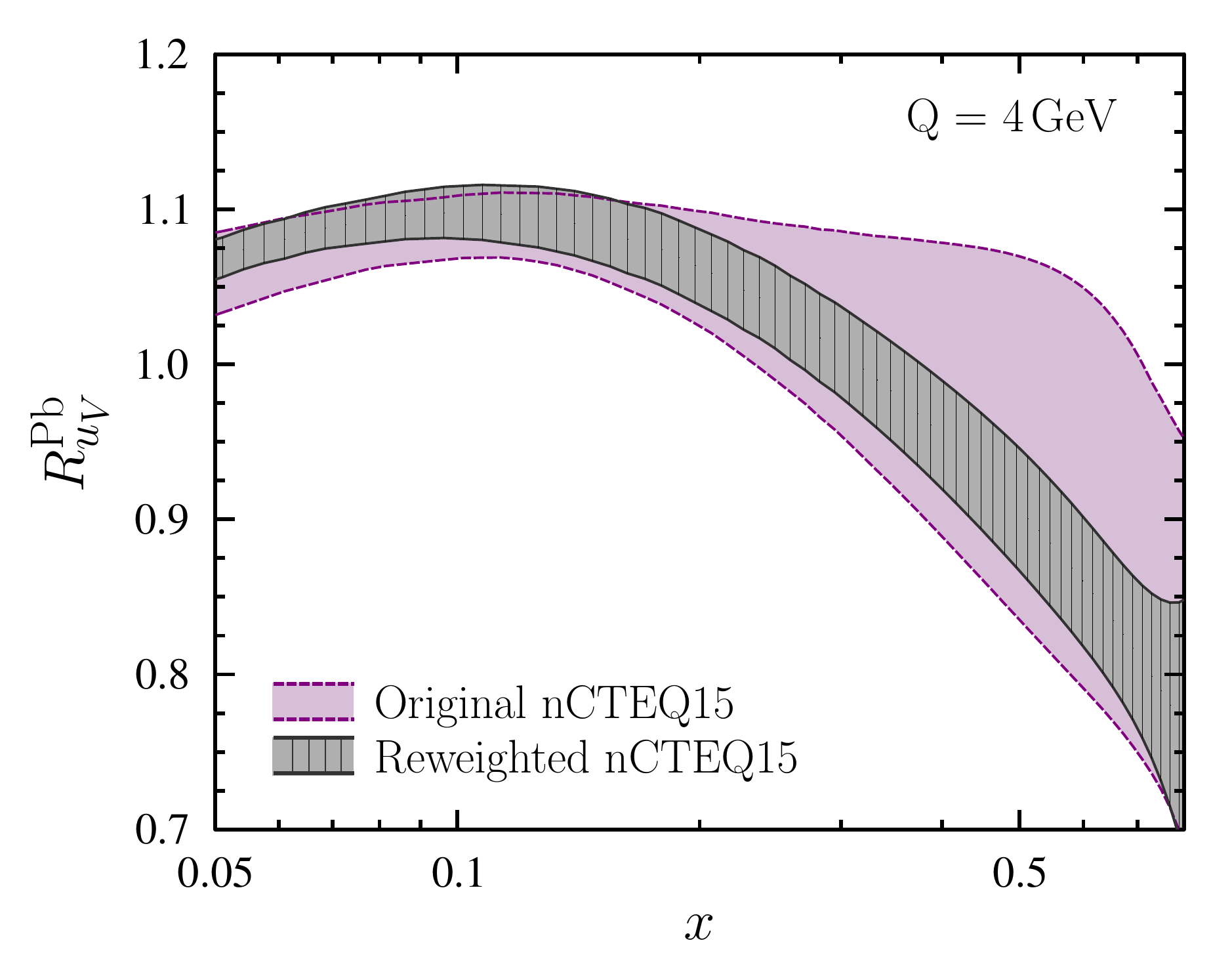}
\includegraphics[width=0.329\linewidth]{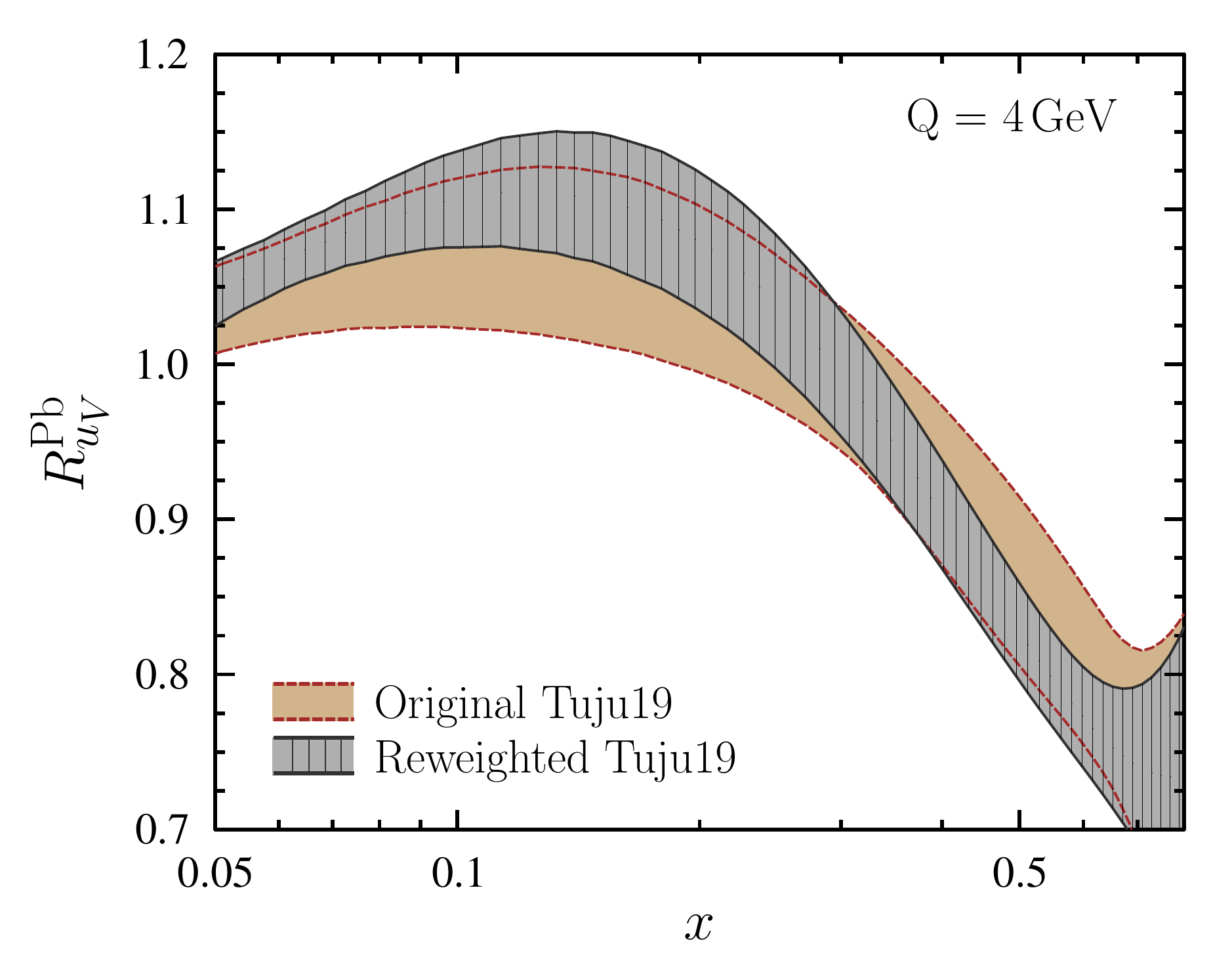}
\caption{Effect of re-weighting on nuclear modifications of down (upper row) and up (lower row) valence quarks in Pb at $Q=4$ GeV. The coloured bands are the original uncertainty bands and the grey bands with hatching are the ones after re-weighting.}
\label{fig:valencerw}
\end{figure*} 

A more quantitative estimate of the data-to-theory correspondence can be obtained by looking at the $\chi^2$ values. To this intent, we computed the $\chi^{2}$ for all the central and error sets. The resulting values are displayed in Fig.~\ref{fig:chi2s}. The central values are $\chi^2/N_{\rm data} = 0.93$ for EPPS16, $\chi^2/N_{\rm data} = 0.98$ for nCTEQ15, and $\chi^2/N_{\rm data} = 4.4$ for TuJu19. Thus, the central sets of EPPS16 and nCTEQ15 are well compatible with the CLAS data while TuJu19 is not. As can be seen from Fig.~\ref{fig:chi2s}, the $\chi^2$ values given by the EPPS16 error sets are all very similar and close to the central value. This insensitivity implies that the CLAS data are well compatible with EPPS16 and that they will not have a very significant effect if included in the analysis. In the case of nCTEQ15 there is clearly much more variation from one error set to another. Indeed, the error sets 24, 26, 30 and 33 evidently stick out from the rest. For the high values of $\chi^2$ these error sets correspond to points in the fit-parameter space that are incompatible with the CLAS data. Given that the nCTEQ15 tolerance $\Delta\chi^{2}= 35$ is much less than the variation we see in Fig.~\ref{fig:chi2s} it can be expected that the CLAS data will have a notable impact on nCTEQ15. There are also quite some variation in the $\chi^2$ values obtained with TuJu19 error sets. While none of the error sets agree with the data, the observed variation implies that it is possible to find combinations of error sets that improve the agreement.  

\begin{table}[htb]
\caption{The values of $\chi^2/N_{\mathrm{data}}$ for the original central PDF set and for the central set after re-weighting analysis. The induced penalties and original tolerance criteria $\Delta\chi^2$ are indicated as well.}
\begin{center}
\begin{tabular}{ccccc}
PDF set & $\frac{\chi^2_{\mathrm{orig.}}}{N_{\mathrm{data}}}$ & $\frac{\chi^2_{\mathrm{rew.}}}{N_{\mathrm{data}}} $ &  PDF penalty & $\Delta\chi^2$ \\
\hline
EPPS16  & 0.93 & 0.78 & 5.9 & 52 \\
nCTEQ15 & 0.98 & 0.72 & 3.9 & 35 \\
Tuju19  & 4.4  & 1.5  & 72  & 50 \\
\hline
\end{tabular}
\end{center}
\label{tab:chi2_penalty}
\end{table}

The results of re-weighting are presented in Figs.~\ref{fig:reweighteddata} and \ref{fig:valencerw}, with some characteristics given in Table~\ref{tab:chi2_penalty}. From the numbers in Table~\ref{tab:chi2_penalty} we see that the re-weighting has been able to decrease the central value of $\chi^2$ by some tens of units in the case of EPPS16/nCTEQ15, and by some staggering 300 units in the case of TuJu19. The estimated increase in the original minimum $\chi^2$ (PDF penalty) in the EPPS16 and nCTEQ15 analyses is only a few units -- clearly less than the tolerances $\Delta\chi^2$. These numbers corroborate the fact that the CLAS data are fully compatible with these two sets of PDFs. In the case of TuJu19 the penalty is $\sim 70$ units which clearly exceeds the estimated error tolerance $\Delta\chi^2_{\rm TuJu19}=50$. Thus, although the new central value $\chi^2/N_{\rm data} = 1.5$ is acceptable, some other data in the TuJu19 analysis are no longer satisfactorily reproduced. This means that there is a striking contradiction between the TuJu19 analysis and the CLAS data.

In Fig.~\ref{fig:reweighteddata} we present a comparison between the original error bands of Fig.~\ref{fig:orig} and the ones after re-weighting the PDFs with the CLAS data. As anticipated, the re-weighting has induced only modest effects on EPPS16 predictions which are barely visible for other than the two heaviest nuclei.  In the case of nCTEQ15 the re-weighted error bands are notably narrower than the original ones -- more than a factor of two in some places. For both EPPS16 and nCTE15 the optimal shifts in the data due to the normalization uncertainties are not particularly large. In the case of TuJu19 the re-weighting has induced a quite significant change in the EMC slope. The partons have adjusted themselves to clearly steepen the originally too flat EMC slope and also the uncertainties are somewhat reduced. Thus, even if there are now incompatibilities between the original TuJu19 fit and the CLAS data, the uncertainties do not generally grow. The optimal shifts in the central data values are also larger than in the case of EPPS16/nCTEQ15. 

The original up- and down-valence distributions of Fig.~\ref{fig:valenceall} are compared with the re-weighted ones in Fig.~\ref{fig:valencerw} for EPPS16 (left panels), nCTEQ15 (middle panels) and TuJu19 (right panels). The upper row corresponds to the valence down-quark distributions which seem to remain rather stable upon performing the re-weighting. The lower panels correspond to the valence up-quark distributions. Again, EPPS16 remains nearly unchanged while there are now significant differences in nCTEQ15 and TuJu19. In the case of nCTEQ15 the theoretical uncertainties for the up-valence distribution reduce quite dramatically in the region spanned by the data. Through the assumed form of the fit functions these improvements are also reflected at smaller $x$. The reason why the CLAS data has restricted particularly the up-valence distributions can be understood on the basis of non-isoscalarity of the heaviest CLAS nucleus.
Indeed, e.g. for Pb nucleus, 
\begin{equation}
\frac{Z  + 4N }{4Z  + N} \approx 1.3 \,,
\end{equation}
so that the CLAS data are sensitive to several different linear combinations of $R_{u_V}^{p/A}$ and $R_{d_V}^{p/A}$ -- not just the one indicated in the last row of Eq.~(\ref{eq:NCDIS}) when only isoscalar nuclei are used in the fit. As a result, one can better unfold both $R_{u_V}^{p/A}$ and $R_{d_V}^{p/A}$ separately. Since $R_{u_V}^{p/A}$ is better constrained already before the re-weighting most of the new constraints go to $R_{d_V}^{p/A}$. From Eqs.~(\ref{eq:Ruvtot2}) and (\ref{eq:Rdvtot2}) we in turn see, that when $R_{d_V}^{p/A}$ gets better constrained the impact is stronger in $R_{u_V}^{\rm Pb}$. This explains the hierarchy seen in Fig.~\ref{fig:valencerw} for nCTEQ15. For TuJu19 the main effect is that the up-valence distribution has become steeper from its original shape. This increased steepness is in agreement with the steeper cross sections observed in Fig.~\ref{fig:reweighteddata}. 

\begin{figure}[htb!]
\centering
\includegraphics[width=1.000\linewidth]{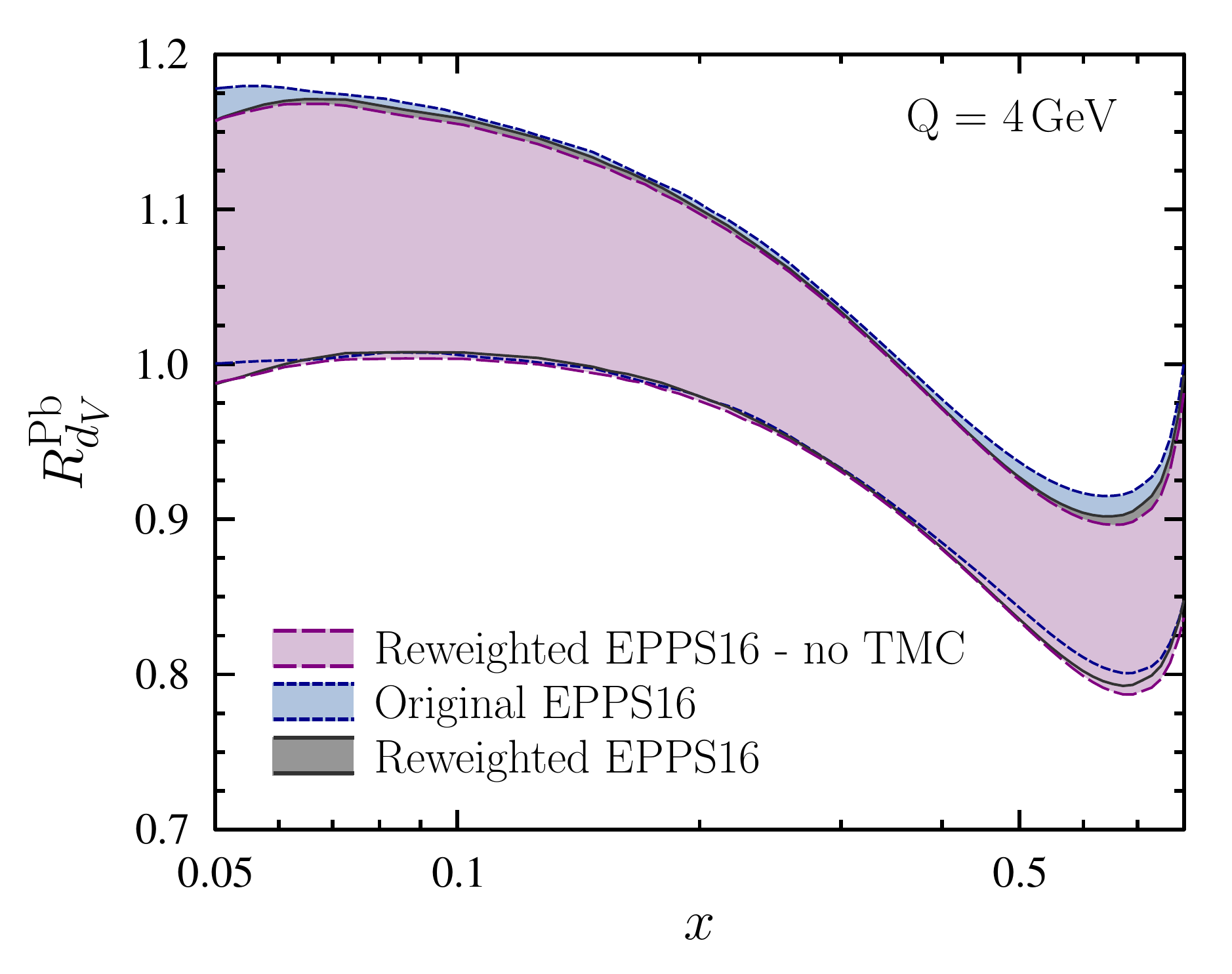}
\includegraphics[width=1.000\linewidth]{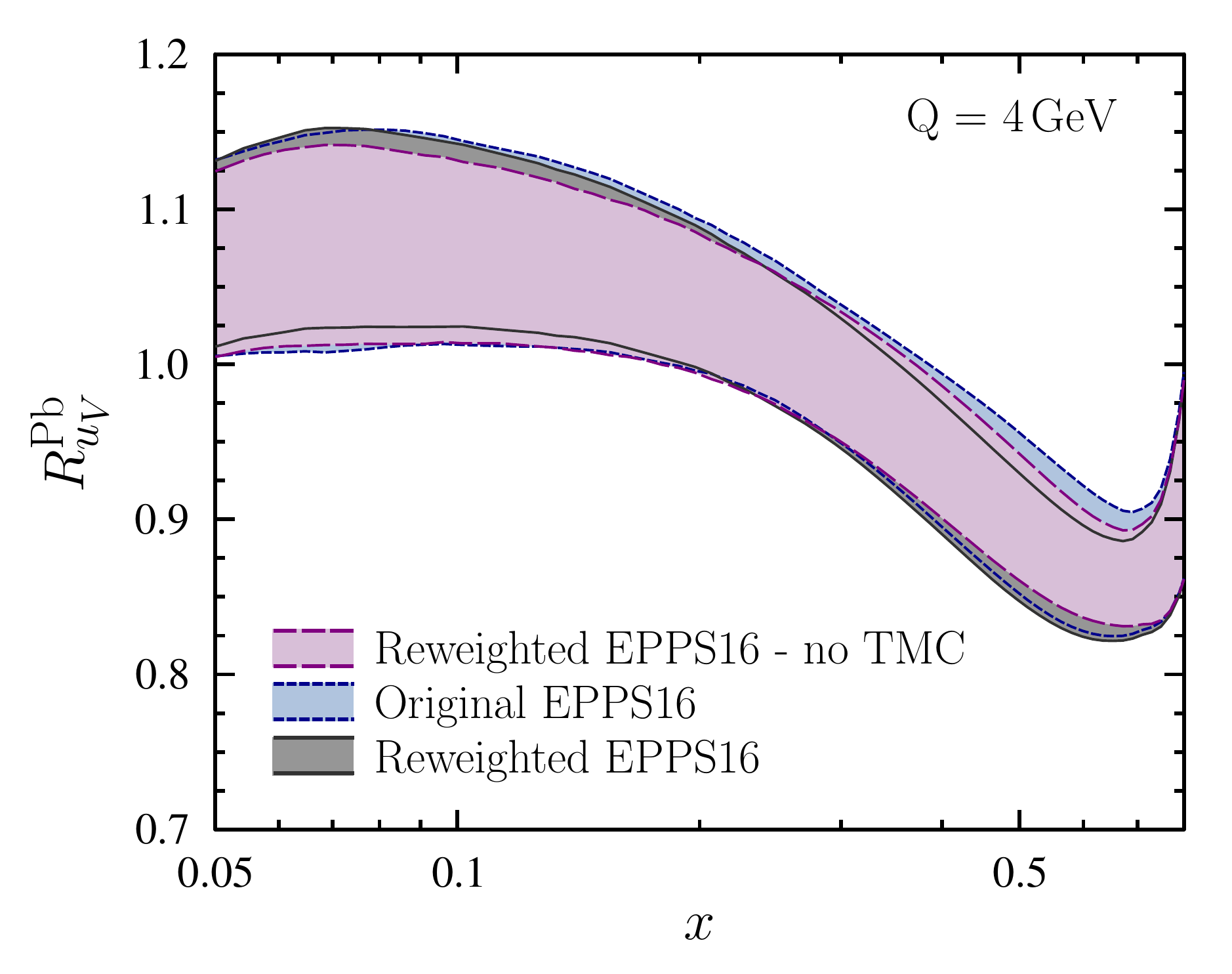} \\
\caption{Effect of re-weighting on EPPS16 with and without the TMCs.}
\label{fig:TMC_vs_noTMC}
\end{figure}

As a final exercise we have investigated the role of TMCs when performing the re-weighting. In Fig.~\ref{fig:TMC_vs_noTMC} we plot the EPPS16 up- and down-valence distributions also in the case the TMCs are not applied (in our default results the TMCs are always incorporated). We see that the differences are very moderate between the TMC and no-TMC cases. This is presumably related to the overall normalization uncertainties which can hide the differences seen in Fig.~\ref{fig:orig} if the TMCs are not applied, by appropriately reshuffling the systematic parameters $s_k$ in the $\chi^2$ function.

\section{Summary}\label{section:summary}

In the present work we have scrutinized the recent high-$x$ neutral-current DIS data measured by the CLAS collaboration. In particular, we have investigated whether these data are in agreement with the modern nuclear PDFs and whether they could provide additional constraints. We have found that the data agree nicely with the EPPS16 and nCTEQ15 global analyses of nuclear PDFs while they disagree with TuJu19. As a feasible explanation for the clash with TuJu19 we entertained the possibility that extending the parametrization of the $A$ dependence down to deuteron may bias the predictions at large $A$. In any case, from the good agreement with EPPS16 and nCTEQ15 we can conclude that these data are compatible with the other world data used in global fits of nuclear PDFs. 

What is also interesting here is that the CLAS data are situated at lower $Q^2$ than the other large-$x$ data in the global fits. The agreement we find indicates that there are no significant additional higher-twist contributions present. Although the target-mass effects are of the same size as the uncorrelated CLAS data uncertainties, their impact in the global analysis is predicted to be small due to the normalization uncertainties that can partly shroud these effects. 
Including TMCs in future fits of nuclear PDFs would then be a recommendable but not a crucial practice.
In addition, the nuclear PDFs do not encode non-trivial nuclear effects that would depend on the isospin. We thus find no evidence of short-range correlations or equivalent phenomena that would depend on the relative number of protons and neutrons in the nuclei. This is in line with the results of e.g. Ref.~\cite{Arrington:2019wky}. Our findings allow us then to give an affirmative answer to the question raised in the title.

\section*{Acknowledgements}

We thank E. Segarra for providing us the CLAS data. P.Z. was partially supported by the Deutsche Forschungsgemeinschaft (DFG, German Research Foundation) - Research Unit FOR 2926, grant number 409651613. H.P. wishes to acknowledge the funding from the Academy of Finland project 308301.

\bibliographystyle{unsrt}
\bibliography{main}

\end{document}